\documentclass[11pt,letterpaper]{article}
\usepackage[T1]{fontenc}
\usepackage[utf8]{inputenc}
\usepackage{lmodern}
\usepackage[margin=1in]{geometry}
\usepackage{amsmath,amssymb,amsfonts}
\usepackage{graphicx}
\usepackage{booktabs,array,tabularx,multirow}
\usepackage{subfig}
\usepackage{placeins}
\usepackage{xcolor}
\usepackage[authoryear,round]{natbib}
\usepackage{url}
\usepackage[hidelinks]{hyperref}
\usepackage{microtype}
\usepackage{setspace}
\newcommand{\ArxivAffiliations}{}
\newcommand{\ArxivContact}{}
\newcommand{\ArxivAbstractText}{}
\newcommand{\ArxivKeywordsText}{}
\newcommand{\Title}[1]{\title{#1}}
\newcommand{\Author}[1]{%
  \author{\parbox{0.96\textwidth}{\centering #1}\\[1.1em]
  \begin{minipage}{0.96\textwidth}\centering\footnotesize
  \ArxivAffiliations\\[0.8em]\ArxivContact
  \end{minipage}}
}
\newcommand{\AuthorNames}[1]{\hypersetup{pdfauthor={#1}}}
\newcommand{\address}[1]{\gdef\ArxivAffiliations{#1}}
\newcommand{\corres}[1]{\gdef\ArxivContact{#1}}
\newcommand{\PaperAbstract}[1]{\gdef\ArxivAbstractText{#1}}
\newcommand{\keyword}[1]{\gdef\ArxivKeywordsText{#1}}
\date{}

\newcommand{\authorcontributions}[1]{\par\medskip\noindent\textbf{Author Contributions:} #1\par}
\newcommand{\funding}[1]{\par\smallskip\noindent\textbf{Funding:} #1\par}
\newcommand{\institutionalreview}[1]{\par\smallskip\noindent\textbf{Institutional Review Board Statement:} #1\par}
\newcommand{\informedconsent}[1]{\par\smallskip\noindent\textbf{Informed Consent Statement:} #1\par}
\newcommand{\dataavailability}[1]{\par\smallskip\noindent\textbf{Data Availability Statement:} #1\par}
\newcommand{\conflictsofinterest}[1]{\par\smallskip\noindent\textbf{Conflicts of Interest:} #1\par}

\newcommand{\appendixstart}{%
  \clearpage
  \setcounter{figure}{0}\setcounter{table}{0}\setcounter{equation}{0}
  \renewcommand{\thefigure}{A\arabic{figure}}
  \renewcommand{\thetable}{A\arabic{table}}
  \renewcommand{\theequation}{A\arabic{equation}}
  \renewcommand{\theHfigure}{appendix.\arabic{figure}}
  \renewcommand{\theHtable}{appendix.\arabic{table}}
  \renewcommand{\theHequation}{appendix.\arabic{equation}}
}
\newlength{\extralength}
\newenvironment{adjustwidth}[2]{}{}
\newcommand{\reftitle}[1]{} 
\newcommand{\PublishersNote}[1]{} 

\AtBeginDocument{%
  \maketitle
  \begin{abstract}\ArxivAbstractText\end{abstract}
  \par\bigskip
}

\Title{Dynamic Scenario-Based Portfolio Optimization for Multi-Asset ETFs}

\Author{Abigail A. Mensah$^{1,}$*,
        Ali M. Jaffri$^{3}$,
        Nicholas Appiah$^{1}$,
        W.\ Brent Lindquist$^{1}$,
        Svetlozar T.\ Rachev$^{1}$ and
        Frank J.\ Fabozzi$^{2}$}

\AuthorNames{Abigail A. Mensah, Ali M. Jaffri, Nicholas Appiah,
             W. Brent Lindquist, Svetlozar T. Rachev and Frank J. Fabozzi}

\address{%
$^{1}$ \quad Department of Mathematics \& Statistics, Texas Tech University,
             Lubbock, TX 79409, USA\\
$^{2}$ \quad Carey Business School, Johns Hopkins University,
             Baltimore, MD 21202, USA\\
$^{3}$ \quad College of Business, North Dakota State University,
             Fargo, ND 58102, USA}

\corres{Correspondence: abigail.a.mensah@ttu.edu}

\PaperAbstract{Portfolio outcomes may depend as much on trading restrictions as on how returns are modeled. We compared historical moving-window returns with conditional scenarios from ARMA--GARCH models and a Student-$t$ copula for 30 U.S.-listed exchange-traded funds. Matched mean--variance and conditional value-at-risk portfolios were evaluated under long-only and limited long--short bounds over 946 sessions. Historical portfolios had higher observed reward--risk ratios in most comparisons, but no paired difference remained significant after multiple-testing adjustment. Conditional scenarios did not consistently reduce realized downside risk. We distinguished changes in target weights from trades required to rebalance holdings after market movements. Relaxing the target-adjustment limit changed exposures, turnover, and some returns more than varying scenario count or seed in the examined tests. These findings suggest that assessments of conditional return inputs should consider portfolio constraints and trading requirements.}

\keyword{exchange-traded funds; portfolio optimization; financial risk management; return scenarios; conditional value at risk; turnover constraints; transaction costs}

\begin{document}



\section{Introduction}
\label{sec:intro}

Exchange-traded funds (ETFs) allow investors to combine equity, fixed-income, real-asset, and alternative exposures. The outcomes of an ETF portfolio depend partly on the returns and risks used to choose the holdings, and partly on the limits placed on changing those holdings. Both considerations matter when market volatility and dependence vary over time \citep{Markowitz1952,Gastineau2010,Cont2001,LonginSolnik2001}.

A historical approach uses the returns observed in a moving estimation window. A conditional approach estimates recent return dynamics and generates possible outcomes for the next period. Conditional scenarios can represent changing volatility and joint losses, but their use also introduces estimation and simulation uncertainty; more detailed return modeling does not necessarily improve realized investment results \citep{Michaud1989,BestGrauer1991,ChopraZiemba1993,DeMiguel2009}. Related JRFM studies consider heavy-tailed commodity ETF portfolios \citep{AppiahEtAl2026} and the role of tail risk and trading costs in predominantly actively managed ETFs \citep{LampteyEtAl2026}. The present study examines a broad multi-asset ETF universe under matched portfolio construction rules, with particular attention to the distinction between desired target weights and the trades required to maintain them.

We compared historical moving-window inputs with conditional scenarios generated from ARMA--GARCH marginal models and a multivariate Student-$t$ copula. Both approaches were applied to the same 30 ETFs using mean--variance and conditional value-at-risk (CVaR) objectives, long-only and limited long--short bounds, equal-weight starting allocations, and the same baseline target-adjustment limit. Realized portfolio returns were calculated from subsequent observed ETF returns, not from the simulated scenarios. Holding the portfolio rules fixed made the two input methods more directly comparable within the stated design, although a restrictive trading rule could also limit the changes either method could make to its holdings.

The paper makes three empirical contributions. First, it compares the realized returns, downside risk, and reward--risk measures of matched historical and conditional portfolios, using paired bootstrap inference to describe uncertainty in their differences. Second, it distinguishes adjustments between successive target weights from the rebalancing trades required after market returns cause holdings to drift. This provides a separate view of portfolio turnover, exposures, and modeled trading costs. Third, it examines whether the observed comparisons change when the daily adjustment limit, simulated scenario count, random seed, or estimation-window length is varied. The paper does not propose a new return model or optimization criterion. Its contribution is the empirical assessment of how return-input construction interacts with the rules governing portfolio holdings and trading.

The evaluation covers 946 trading sessions from December 2022 to September 2026. Execution prices, market impact, and short-position financing were not observed, so reported performance follows a close-to-close return-attribution convention with illustrative proportional costs. Section~\ref{sec:data} describes the data, Section~\ref{sec:construction} presents the portfolio methods, Section~\ref{sec:results} reports the empirical results, and Sections~\ref{sec:discussion} and \ref{sec:conclusions} discuss their implications and limitations.

\section{Data and Portfolio Design}
\label{sec:data}

The analysis included 30 U.S.-listed ETFs spanning equities, fixed income, income-oriented and multi-asset funds, real assets, and alternative strategies (Appendix Table~\ref{tab:ETFList}). This selection provided different sources of market exposure, although several funds held overlapping underlying assets. The resulting portfolios were therefore diversified across ETF tickers, not necessarily across independent economic risks.

The adjusted closing prices of the 30 U.S.-listed ETFs were downloaded from Yahoo Finance on 27~October~2026 using Python's \texttt{yfinance} interface. The download requested daily observations from 19~December~2018 (inclusive) to 1~October~2026 (exclusive), with \texttt{auto\_adjust=True}, \texttt{actions=True}, \texttt{threads=False}, and \texttt{progress=False}. Under the specified adjustment option, the returned \texttt{Close} field contains Yahoo Finance's adjusted closing prices, incorporating its treatment of distributions and splits. The corporate-action fields were requested but not separately archived in the analysis workbook; the provider's adjustment factors were not independently reconstructed.

The retrieved price columns were ordered to match the fixed 30-ETF universe. The common panel was formed by deleting dates missing a price for any ETF, using \texttt{dropna()}; no interpolation, forward filling, or other imputation was applied. Examination of the archived workbook found that its \texttt{Full\_Price\_Panel} and \texttt{Adjusted\_Prices} sheets were identical and contained no missing or nonpositive prices. The panel comprised 1,955 price observations from 19~December~2018 to 30~September~2026 and 1,954 one-session arithmetic returns. Its trading dates were compared with the New York Stock Exchange session calendar; no missing, extra, or duplicated sessions were identified. Recalculating every arithmetic return from successive adjusted prices agreed with the saved return sheet to numerical precision.

As a limited check of corporate-action treatment, issuer and exchange records were examined for the one-for-three reverse split of SRET effective 20~December~2022, the BYLD distribution dated 2~September~2026, and the IYG distribution dated 15~June~2026. These events fell on dates present in the archived panel, and the adjusted SRET series showed no mechanical threefold jump across the split. The checks did not independently validate every historical dividend or split factor: raw unadjusted closing prices and corporate-action fields were not saved with this workbook. The study therefore used the provider-adjusted return convention rather than an independently reconstructed total-return index.

For ETF $i$, the daily arithmetic return was calculated as
\begin{equation}
r_{i,t}=\frac{P_{i,t}}{P_{i,t-1}}-1,
\end{equation}
where $P_{i,t}$ denotes the provider-adjusted closing price for ETF $i$ at session $t$. The return series therefore reflects the provider's adjustment convention, rather than an independently reconstructed distribution-reinvested index.

The baseline estimation window contained $W=1{,}008$ returns and ended initially on 20~December~2022. The first evaluated portfolio return was dated 21~December~2022, and 946 out-of-sample sessions were evaluated through 30~September~2026. Both optimization methods and all reported portfolio performance calculations used the same synchronized return panel; the indexed-price figures used the corresponding adjusted prices. Using a common panel removes differences in underlying asset-return samples as an explanation for the compared portfolio histories. The historical distribution-adjustment factors supplied by the provider were not independently reconstructed.

The U.S. three-month Treasury par yield was the risk-free proxy. At each decision date, the most recent available yield was converted to an effective daily rate under a 252-session convention (Equation~\ref{eq:riskfree_conversion}). The aligned rates were used both in estimated excess-return objectives and in realized excess-return ratios.

The benchmark was an initially equal-weighted buy-and-hold portfolio (BHP), with $1/30$ of capital allocated to each ETF and no subsequent rebalancing. It provides a common reference for realized wealth, although it does not maintain the target exposures of an optimized strategy. All wealth paths began at USD~100. The return-correlation matrix and indexed adjusted-price series in Appendix Figures~\ref{fig:etf_correlation_2026} and \ref{fig:etf_indexed_prices_2026} describe the ETF universe rather than portfolio investment outcomes.

\section{Portfolio Construction}
\label{sec:construction}

At each decision date, both methods used information from the same moving window of $W=1{,}008$ daily returns for $N=30$ ETFs. The objectives, position bounds, starting weights, and baseline target-adjustment cap were held fixed. Only the representation of returns supplied to the optimizer differed.

Let \(\boldsymbol{r}_t=(r_{1,t},\ldots,r_{N,t})^{\mathsf T}\)
denote the vector of arithmetic ETF returns at trading
date \(t\). The historical approach used the
empirical return matrix
\begin{equation}
\boldsymbol{X}_{t}^{H}
=
\begin{pmatrix}
\boldsymbol{r}_{t-W+1}^{\mathsf T}\\
\vdots\\
\boldsymbol{r}_{t}^{\mathsf T}
\end{pmatrix}
\in\mathbb{R}^{W\times N}.
\end{equation}
The observations in this matrix were supplied directly
to the mean--variance and conditional value at risk
(CVaR) optimizers.

The dynamic approach followed the scenario-based framework
described by \citet[Chapter~7]{Lindquist2022} and
\citet{Jaffri2025}. For each ETF, an ARMA(1,1)--GARCH(1,1)
model was estimated using the current moving window \citep{Engle1982,Bollerslev1986}.
The conditional mean equation was specified as
\begin{equation}
r_{i,\ell}
=
\delta_{0,i}
+
\delta_{1,i}r_{i,\ell-1}
+
\theta_i\varepsilon_{i,\ell-1}
+
\varepsilon_{i,\ell},
\end{equation}
where
\begin{equation}
\varepsilon_{i,\ell}
=
\sigma_{i,\ell}z_{i,\ell},
\end{equation}
and the conditional variance followed
\begin{equation}
\sigma_{i,\ell}^{2}
=
\omega_i
+
\alpha_i\varepsilon_{i,\ell-1}^{2}
+
\beta_i\sigma_{i,\ell-1}^{2}.
\end{equation}
Here, \(\varepsilon_{i,\ell}\) denotes the return
innovation, \(\sigma_{i,\ell}^{2}\) is its conditional
variance, and \(z_{i,\ell}\) is the standardized
innovation. The model parameters were re-estimated
for each ETF as the moving window advanced.

The marginal models were estimated by maximum likelihood
using a Student-\(t\) innovation specification. However,
the predictive innovations were not generated directly
from the fitted parametric Student-\(t\) distributions.
Instead, the implementation used kernel-based empirical
distributions of the standardized residuals. Thus,
Student-\(t\) innovations characterized the estimation
likelihood, while empirical residual distributions
provided the marginal simulation inputs.

The standardized residuals were calculated as
\begin{equation}
\widehat z_{i,\ell}
=
\frac{\widehat\varepsilon_{i,\ell}}
{\widehat\sigma_{i,\ell}}.
\end{equation}
For each ETF, a kernel-based cumulative distribution
estimate \(\widehat F_{i,t}\) was constructed from
the standardized residuals in the current window.
The residual observations were transformed to the
unit interval using
\begin{equation}
u_{i,\ell}
=
\widehat F_{i,t}
\left(\widehat z_{i,\ell}\right).
\end{equation}
Before copula fitting, the implementation restricted the kernel-CDF
pseudo-observations to the open unit interval according to
\begin{equation}
u_{i,\ell}^{*}=\min\left\{1-\epsilon,
\max\left(\epsilon,u_{i,\ell}\right)\right\},
\qquad \epsilon=10^{-6}.
\end{equation}
The same clipping was applied to the uniforms generated from the
fitted copula before kernel-inverse simulation. The transform used
a kernel CDF rather than rank-based pseudo-observations.

A multivariate Student-\(t\) copula was fitted to
the resulting \(W\times N\) matrix of transformed
standardized residuals using MATLAB's approximate
maximum-likelihood procedure. The fitted copula
represented cross-sectional dependence among ETF
innovations, including the symmetric tail dependence
admitted by the Student-\(t\) copula family
\citep{McNeil2005,Patton2006}. The dependence parameters were re-estimated at each portfolio
decision. In particular, MATLAB's
\texttt{copulafit} with the Student-$t$ family and approximate maximum likelihood
returns a copula correlation matrix and fitted degrees of freedom,
and \texttt{copularnd} draws uniforms using those fitted parameters.
The model-estimation procedure recorded fitted copula degrees of freedom, selected correlation and eigenvalue diagnostics, kernel bandwidths, standardized-residual summaries, and numerical warnings. These diagnostics documented model estimation but did not independently validate future tail outcomes.

For each moving window, \(S=10{,}000\) joint
pseudo-uniform scenario vectors were generated
from the fitted copula. The simulated standardized
innovations were obtained by applying the
kernel-based empirical inverse distributions:
\begin{equation}
z_{i,s,t+1}
=
\widehat F_{i,t}^{-1}
\left(u_{i,s,t+1}\right),
\qquad
s=1,\ldots,S.
\end{equation}
The one-step-ahead conditional mean forecast was
\begin{equation}
\widehat\mu_{i,t+1}
=
\widehat\delta_{0,i}
+
\widehat\delta_{1,i}r_{i,t}
+
\widehat\theta_i\widehat\varepsilon_{i,t},
\end{equation}
and the corresponding variance forecast was
\begin{equation}
\widehat\sigma_{i,t+1}^{2}
=
\widehat\omega_i
+
\widehat\alpha_i\widehat\varepsilon_{i,t}^{2}
+
\widehat\beta_i\widehat\sigma_{i,t}^{2}.
\end{equation}
The implemented one-step-ahead scenario-return specification is
\begin{equation}
r_{i,s,t+1}^{D}
=
\widehat\mu_{i,t+1}
+
\widehat\sigma_{i,t+1}z_{i,s,t+1}.
\end{equation}

The resulting scenario matrix,
\begin{equation}
\boldsymbol{X}_{t}^{D}
=
\begin{pmatrix}
(\boldsymbol{r}_{1,t+1}^{D})^{\mathsf T}\\
\vdots\\
(\boldsymbol{r}_{S,t+1}^{D})^{\mathsf T}
\end{pmatrix}
\in\mathbb{R}^{S\times N},
\end{equation}
was used as the return input for dynamic portfolio
optimization. The historical procedure used the observed return vectors, while the conditional procedure used simulated vectors drawn from estimates based on those same observations. Simulation enlarged the number of optimization inputs but did not add market observations. Its potential value lay in changing how volatility and joint losses were represented, not in enlarging the underlying information set.

For either approach, let
\(\widehat{\boldsymbol\mu}_t\) and
\(\widehat{\boldsymbol\Sigma}_t\) denote the
sample mean vector and covariance matrix of
the corresponding optimization inputs.
Mean--variance optimization followed
\citet{Markowitz1952}. The minimum-variance
portfolio was defined by
\begin{equation}
\begin{aligned}
\min_{\boldsymbol w}\quad&
\boldsymbol w^{\mathsf T}
\widehat{\boldsymbol\Sigma}_t
\boldsymbol w\\
\text{subject to}\quad&
\boldsymbol{1}^{\mathsf T}\boldsymbol w=1,\\
&
\boldsymbol\ell\leq\boldsymbol w\leq\boldsymbol u,\\
&
\frac12
\left\|
\boldsymbol w-\boldsymbol w_{t-1}
\right\|_1
\leq \delta.
\end{aligned}
\end{equation}
The associated efficient frontier was obtained
by considering feasible target-return conditions,
\begin{equation}
\widehat{\boldsymbol\mu}_t^{\mathsf T}
\boldsymbol w
\geq \mu_*,
\end{equation}
under the same investment constraints.

The tangent mean--variance portfolio was selected
according to the excess-return-to-standard-deviation
criterion
\begin{equation}
\operatorname{SR}_t(\boldsymbol w)
=
\frac{
\widehat{\boldsymbol\mu}_t^{\mathsf T}
\boldsymbol w-r_{f,t}
}{
\sqrt{
\boldsymbol w^{\mathsf T}
\widehat{\boldsymbol\Sigma}_t
\boldsymbol w
}
}.
\end{equation}
Here, \(r_{f,t}\) denotes the daily risk-free rate
available at the portfolio decision date.

CVaR portfolios were constructed at confidence levels
\(\alpha\in\{0.95,0.99\}\), following the
sample-average approximation framework of
\citet{RockafellarUryasev2000} and
\citet{KrokhmalPalmquistUryasev2002}.
For portfolio weights \(\boldsymbol w\), the
loss in scenario \(s\) was defined as
\begin{equation}
L_{s,t}(\boldsymbol w)
=
-\boldsymbol w^{\mathsf T}
\boldsymbol x_{s,t},
\end{equation}
where \(\boldsymbol x_{s,t}\) is an empirical
return vector under historical optimization
or a simulated return vector under dynamic
optimization.

Let \(M_t=W\) for historical optimization and
\(M_t=S\) for dynamic optimization. The empirical
CVaR measure was
\begin{equation}
\widehat{\operatorname{CVaR}}_{\alpha,t}
(\boldsymbol w)
=
\min_{\zeta}
\left\{
\zeta+
\frac{1}{(1-\alpha)M_t}
\sum_{s=1}^{M_t}
\left[
L_{s,t}(\boldsymbol w)-\zeta
\right]^+
\right\},
\end{equation}
where \([x]^+=\max(x,0)\). The minimum-CVaR
portfolio minimized this objective subject
to the budget, individual weight bounds,
and target-turnover restriction. At the 99\% confidence level,
the 1,008-return historical window contained approximately 10 observations
in the upper 1\% loss tail, whereas 10,000 conditional scenarios supplied
approximately 100 simulated tail observations. The latter were generated
from estimates based on the same historical window, not from additional
observed market data.

The tangent CVaR portfolio was selected by
comparing the excess-return-to-CVaR criterion
\begin{equation}
\operatorname{TC}_{\alpha,t}(\boldsymbol w)
=
\frac{
\widehat{\boldsymbol\mu}_t^{\mathsf T}
\boldsymbol w-r_{f,t}
}{
\widehat{\operatorname{CVaR}}_{\alpha,t}
(\boldsymbol w)
}
\end{equation}
across the computed frontier portfolios
with finite, strictly positive risk
denominators. Both methods used 20 requested frontier candidates for tangent-CVaR selection. The resulting tangent-CVaR selections were finite-grid approximations rather than solutions to an analytically continuous tangency problem. The 100-point mean--variance frontiers displayed in the illustrative figures excluded turnover constraints and therefore did not measure the resolution of the turnover-constrained tangent-CVaR rule.

We considered long--only (LO) and long--short
(LS) investment specifications. The LO
strategy imposed
\begin{equation}
0\leq w_{i,t}\leq1,
\qquad i=1,\ldots,N,
\end{equation}
while the LS strategy imposed
\begin{equation}
-\frac1N
\leq w_{i,t}
\leq1+\frac1N,
\qquad i=1,\ldots,N.
\end{equation}
Both strategies satisfied
\begin{equation}
\sum_{i=1}^{N}w_{i,t}=1.
\end{equation}
For \(N=30\), the LS specification permitted
each individual short position to have a
weight of at most approximately \(3.33\%\)
in magnitude. The bounds did not impose
a separate portfolio-level gross-exposure
restriction and therefore did not guarantee
low aggregate leverage
\citep{JagannathanMa2003}.

Let $\boldsymbol w_t$ denote the target vector selected using information through date $t$. Its one-way adjustment from the previous target is
\begin{equation}
\tau_t=\frac{1}{2}\sum_{i=1}^{N}|w_{i,t}-w_{i,t-1}|.
\end{equation}
The baseline cap was $\delta=1/(30\times252)$ per decision. It limited changes in target weights rather than a portfolio's gross trading activity. All strategies started from the equal-weight vector \(\boldsymbol w_0=(1/30,\ldots,1/30)\). A sensitivity analysis also considered a 1\% daily target-adjustment allowance to examine how the restriction affected allocations and realized performance.

\paragraph{Computational implementation and numerical validation.}
The 24 portfolio strategies were computed in MATLAB R2026a Update~3. Mean--variance portfolios used the \texttt{Portfolio} class and \texttt{quadprog} with $10^{-9}$ optimality and constraint tolerances; the tangent mean--variance selection used the toolbox's iterative Sharpe-ratio routine. CVaR optimization used \texttt{PortfolioCVaR} and \texttt{TrustRegionCP}, with up to 2,000 iterations and absolute and relative gap tolerances of $10^{-8}$. The tangent-CVaR rule selected the highest admissible ratio from 20 requested frontier candidates. Both input methods used the same bounds, the same cap on changes between target weights, and the same numerical acceptance tests. Numerical cap overshoots no greater than $10^{-6}$ were scaled back toward the preceding feasible target; the cap itself was never enlarged. A failed final candidate would have led to retention of the preceding target.

The saved-output audit found that the final portfolio allocations met the specified budget, position-bound, and target-adjustment checks. The recorded scale-back counts refer to decisions for which at least one optimization candidate required a small numerical correction; they do not necessarily count corrected final allocations. Representative tangent-CVaR comparisons using finer frontier grids and tighter solver settings produced only small differences in selected historical allocations and no recorded differences in the tested conditional allocations. These checks support feasibility and stability for the decisions examined, but neither the original candidate-level target-return constraints nor global optimality was independently verified. Appendix~\ref{app:numerical_checks} reports the correction magnitudes, solver settings, and numerical comparisons.

The conditional simulation used Student-$t$ estimation likelihoods for the ARMA--GARCH marginal models, kernel estimates of their standardized residual distributions, and a fitted Student-$t$ copula. The random seed was fixed before the moving-window loop. Because the simulated residual innovations were not recentered or rescaled, the fitted one-step forecasts need not equal the exact moments of the finite simulated return matrix. The saved diagnostics did not record any reuse of prior marginal fits. The detailed settings and recorded innovation moments are reported in Appendix~\ref{app:numerical_checks}.

To distinguish target adjustments from trades, let $\widetilde{\boldsymbol w}_t$ denote the holdings immediately before the target selected at date $t$ is implemented. The previous target, $\boldsymbol w_{t-1}$, is first allowed to drift with the realized ETF return vector $\boldsymbol r_t$. For ETF $i$, the pre-trade weight is
\begin{equation}
\widetilde w_{i,t}
=\frac{w_{i,t-1}(1+r_{i,t})}
{1+\boldsymbol w_{t-1}^{\mathsf T}\boldsymbol r_t},
\qquad i=1,\ldots,N,
\label{eq:drifted_holdings}
\end{equation}
where the denominator is the gross portfolio wealth factor and is required to be positive. The one-way implementation turnover needed to move from drifted holdings to the newly selected target is
\begin{equation}
\kappa_t=\frac12\sum_{i=1}^{N}
\left|w_{i,t}-\widetilde w_{i,t}\right|.
\label{eq:implementation_turnover}
\end{equation}
The baseline cap applied to target adjustments $\tau_t$, not to implementation turnover $\kappa_t$. In the saved output workbooks, an allocation selected using information through date $t$ was dated by the next trading session, $t+1$, whose return it earned. Accordingly, the turnover recorded under that output date represented $\kappa_t$, calculated after the return on date $t$ and before the next attributed return. The first evaluated allocation had no observed preceding portfolio holdings within the evaluation sample, so its initial formation cost was excluded. These conventions did not assume that actual trades took place at the closing prices used for portfolio estimation.

The U.S.\ three-month Treasury
par yield was used as the
risk-free-rate proxy. Let
\(y_t\) denote its annual
percentage quotation.
The corresponding effective
daily rate was calculated as
\begin{equation}
r_{f,t}
=
\left(
1+\frac{y_t}{100}
\right)^{1/252}-1.
\label{eq:riskfree_conversion}
\end{equation}
The Treasury series was aligned
to the ETF return observation
dates using the most recently
available quotation on or before
each relevant date. The daily
rate available at the end of
the estimation window was used
for portfolio selection, while
date-aligned daily rates were
used to evaluate realized
excess returns. The marginal ARMA--GARCH models were fitted to
raw arithmetic ETF returns, not risk-free-rate-adjusted returns;
therefore, the simulated asset returns are already on the gross
asset-return scale and require no risk-free-rate add-back for
portfolio wealth accounting.

The return panel contained
\(T=1{,}954\) observations.
The first estimation window contained observations 1 through 1,008 and ended on 20~December~2022. The next observed return, dated 21~December~2022, was the first evaluated portfolio return. The number of evaluated observations was
\begin{equation}
T-W=1{,}954-1{,}008=946.
\end{equation}
The evaluation period ends on 30~September~2026. A window without an available next return was not assigned an out-of-sample portfolio return.

At each trading date, the moving window advanced by one observed return. The historical
and dynamic approaches
updated their portfolio
decisions at the same
frequency, with a one-observation
holding period. The dynamic
implementation used
\(\mathrm{HOLD}=1\).
No subsequent realized
return observation entered
the corresponding portfolio
optimization problem.

Let \(\mathcal F_t\) denote
the information available
through the observed return
date \(t\). Portfolio
parameters, predictive
scenarios, and target
allocations were calculated
from this information set.
For weights $\boldsymbol w_t$, the subsequent portfolio return before trading costs was computed as
\begin{equation}
r_{p,t+1}
=
\sum_{i=1}^{N}
w_{i,t}r_{i,t+1}.
\label{eq:gross_next_return}
\end{equation}
The historical and dynamic
strategies were evaluated
against the same realized
ETF returns. Model-generated
scenarios were used for
dynamic portfolio selection
and conditional risk
estimation, not as realized
portfolio returns.

The evaluation employed
a close-to-close return
accounting convention.
Although each target was
computed using information
through the preceding
retained observation,
actual trade execution
times and prices were
not modeled. In particular,
the convention does not
establish that targets
estimated after observing
closing prices could be
executed at those same
closing prices. Consequently,
the reported gross returns
describe the specified
weight-attribution convention
rather than a fully executable
trading strategy.

For comparison across
portfolio objectives,
mean--variance optimization
is denoted by V, while
CVaR optimization at the
95\% and 99\% confidence
levels is denoted by C95
and C99. Minimum-risk
and tangent solutions
are denoted by M and T,
respectively. The resulting
portfolios are labeled
MV, TV, MC95, TC95,
MC99, and TC99, with
an LO or LS prefix
identifying the investment
specification. For example,
LO MC95 denotes the
long--only minimum-CVaR
portfolio at the 95\%
confidence level, whereas
LS TC99 denotes the
long--short tangent-CVaR
portfolio at the 99\%
confidence level.

\section{Empirical Results}
\label{sec:results}
\begin{figure}[htbp]
\centering
\includegraphics[width=\textwidth]{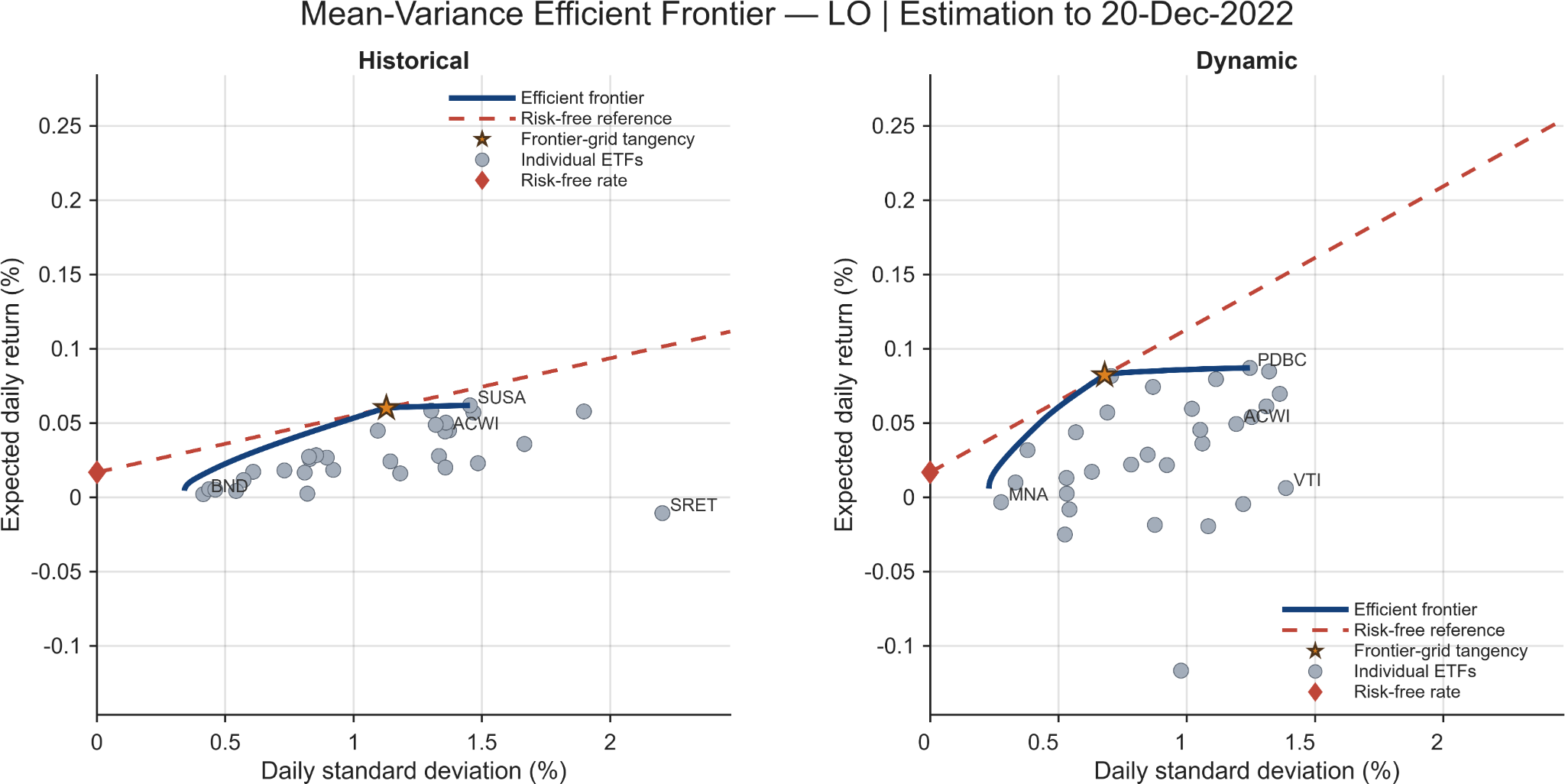}
\caption{Illustrative long--only historical (left) and dynamic (right) mean--variance efficient frontiers. Estimation ends 20~December~2022; evaluated return date 21~December~2022. Return and standard deviation are daily percentages. The plots exclude the baseline turnover restriction. The highlighted point is a frontier-grid maximum excess-return-to-standard-deviation candidate, not the turnover-constrained tangent portfolio.}
\label{fig:frontier_mv}
\end{figure}

\begin{figure}[htbp]
\centering
\includegraphics[width=\textwidth]{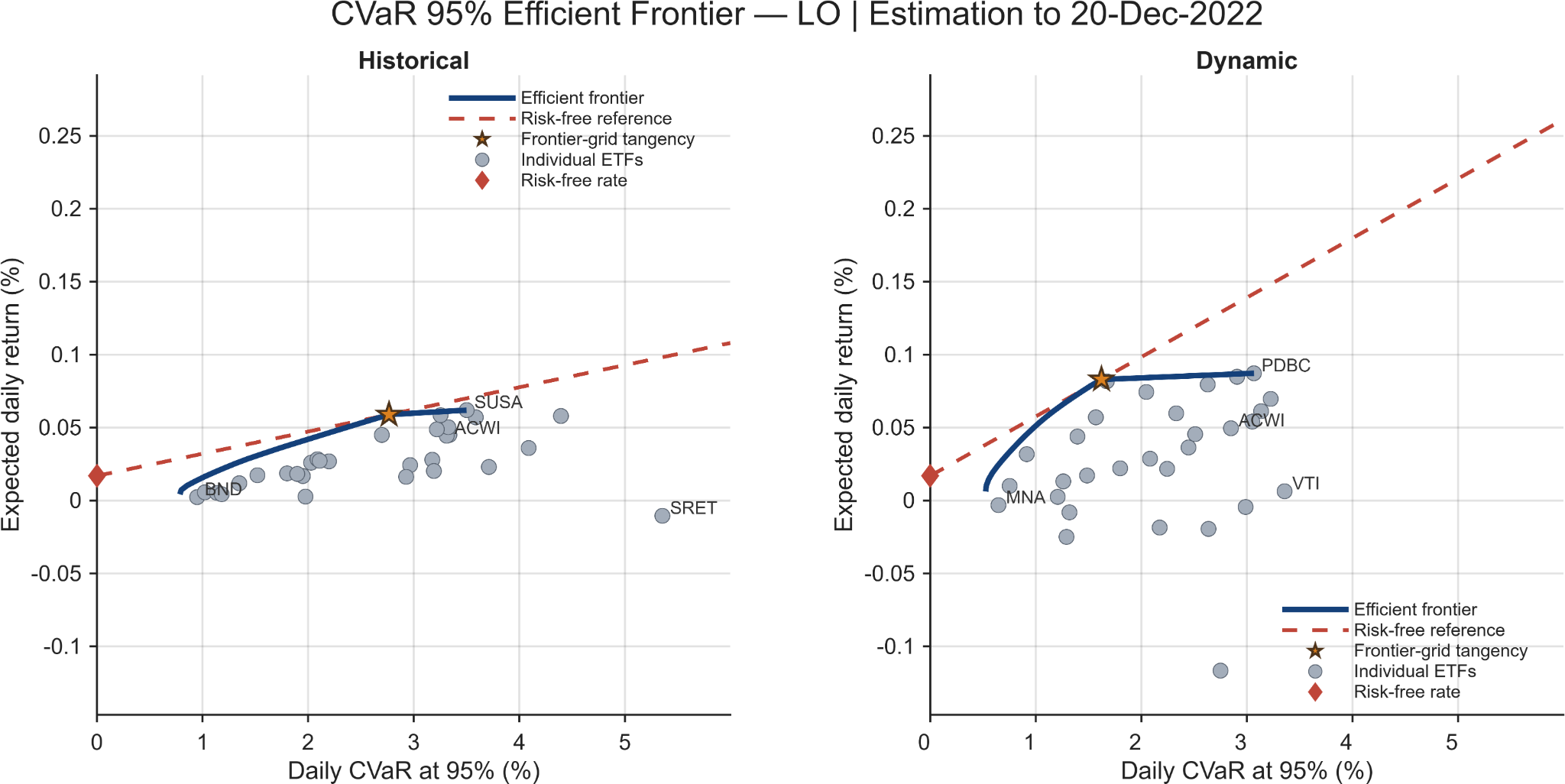}
\caption{Illustrative long--only historical (left) and dynamic (right) CVaR 95\% efficient frontiers at the same estimation cutoff and with identical plotting scales. The plotted candidate maximizes excess expected return divided by CVaR 95\% over the grid; no target-turnover constraint is imposed.}
\label{fig:frontier_cvar95}
\end{figure}

\begin{figure}[htbp]
\centering
\includegraphics[width=\textwidth]{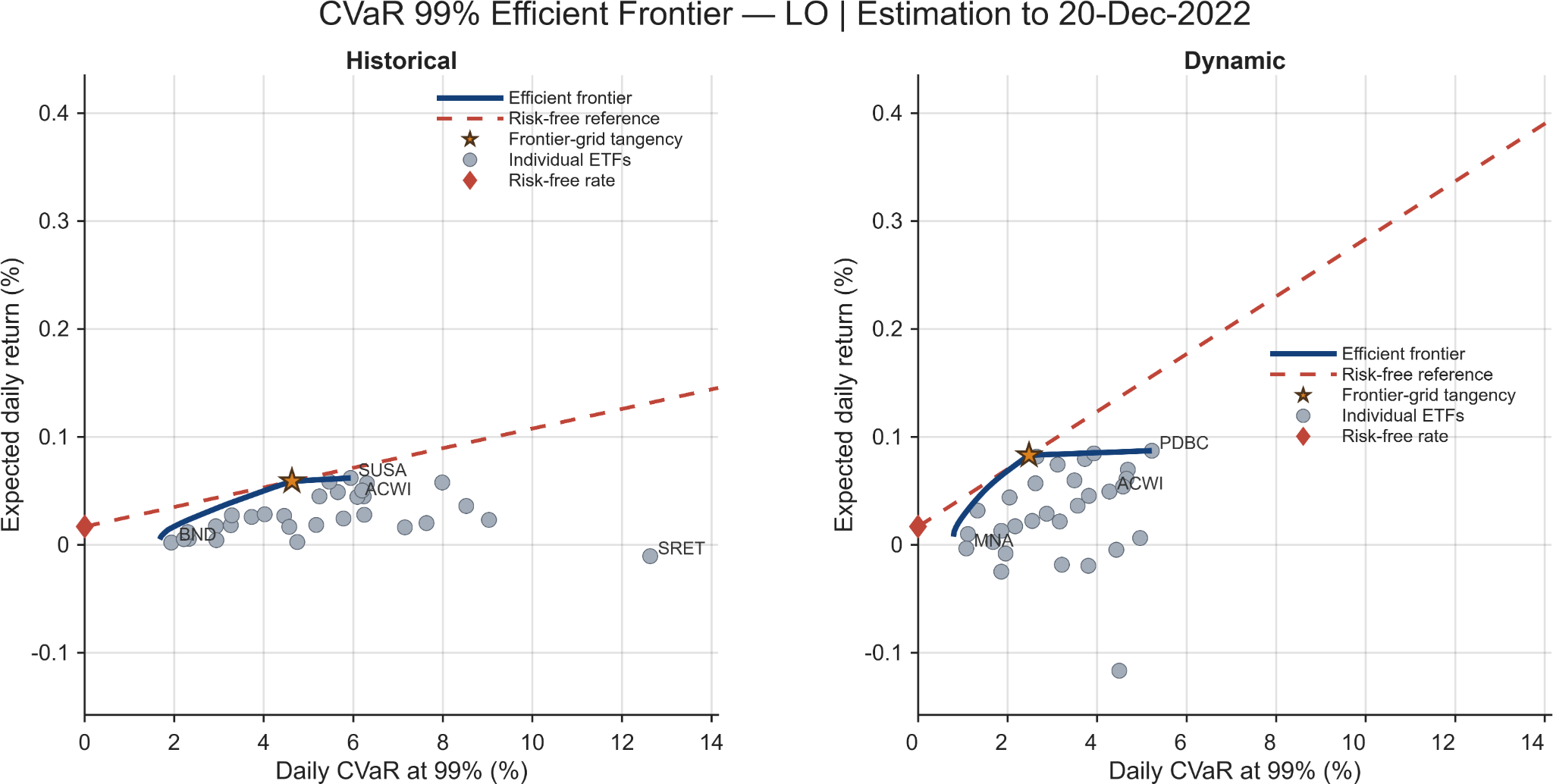}
\caption{Illustrative long--only historical (left) and dynamic (right) CVaR 99\% efficient frontiers. The x-axis is daily CVaR 99\% in percent. The risk-free reference is an illustrative line through the selected excess-return-to-CVaR candidate, not a market-based capital-market line.}
\label{fig:frontier_cvar99}
\end{figure}

The long-only frontiers in Figures~\ref{fig:frontier_mv}--\ref{fig:frontier_cvar99} illustrate how return-input construction affected estimated choices at the first portfolio decision. Both methods used information through 20~December~2022, but the historical frontiers used the observed returns and the conditional frontiers used scenarios generated from the fitted models. Because the panels use matched scales, their differences indicate that the two procedures did not supply identical estimated opportunity sets. These figures exclude the baseline target-adjustment cap, however, and should not be interpreted as the actual constrained allocations or as realized performance comparisons. In the CVaR plots, the dashed reference line illustrates the selected excess-return-to-tail-risk ratio rather than a tradable capital-market line.

The positions of the historical and conditional frontiers differed in Figures~\ref{fig:frontier_mv}--\ref{fig:frontier_cvar99}. The estimated return--risk tradeoffs therefore depended on whether the optimizer received observed returns or model-generated scenarios. These plots describe the respective estimated opportunity sets before the target-adjustment restriction was imposed. They do not show which return forecasts were more accurate or which constrained portfolios subsequently performed better.

\begin{figure}[p]
\centering
\includegraphics[width=0.47\textwidth]{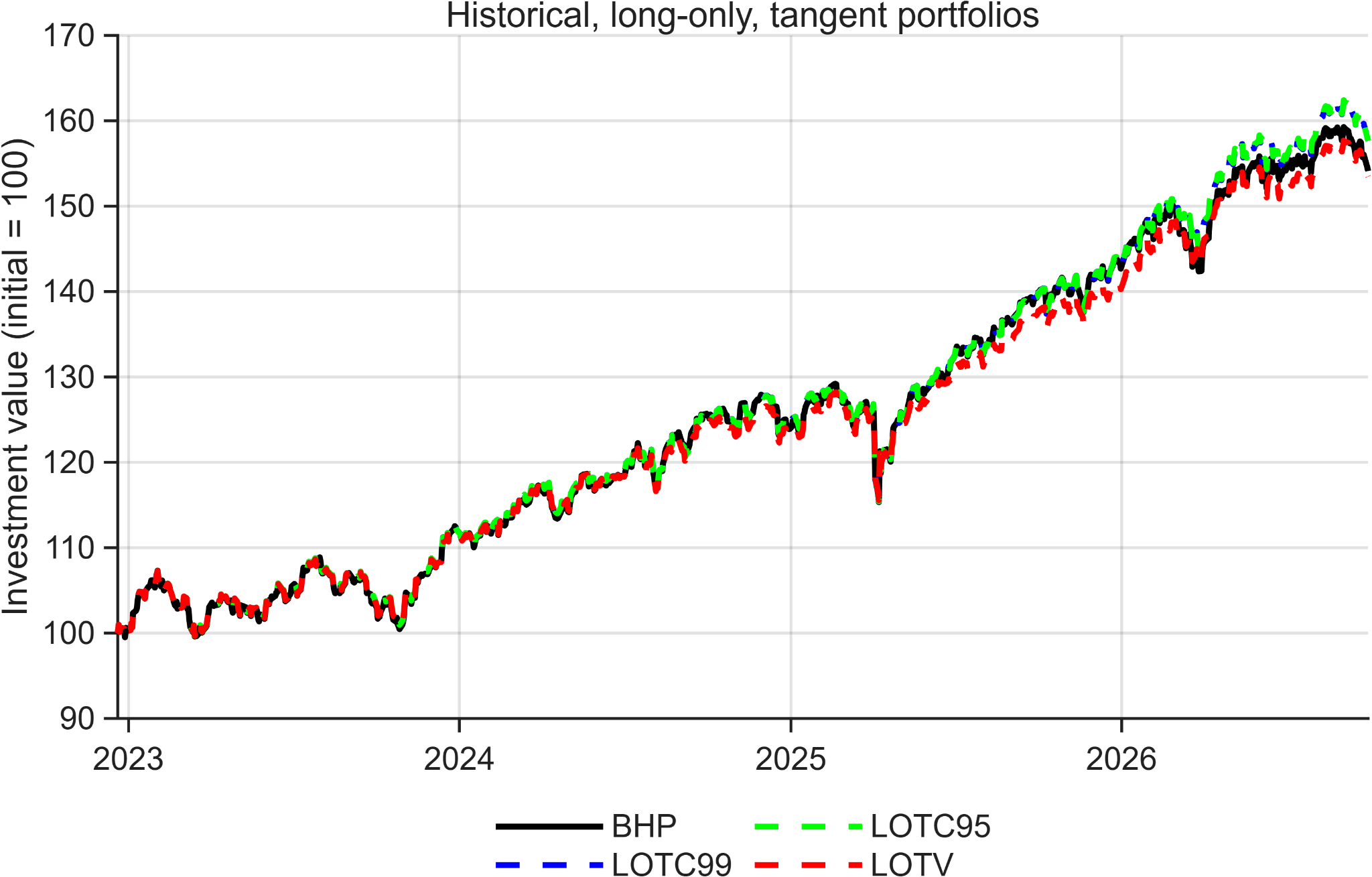}\hfill
\includegraphics[width=0.47\textwidth]{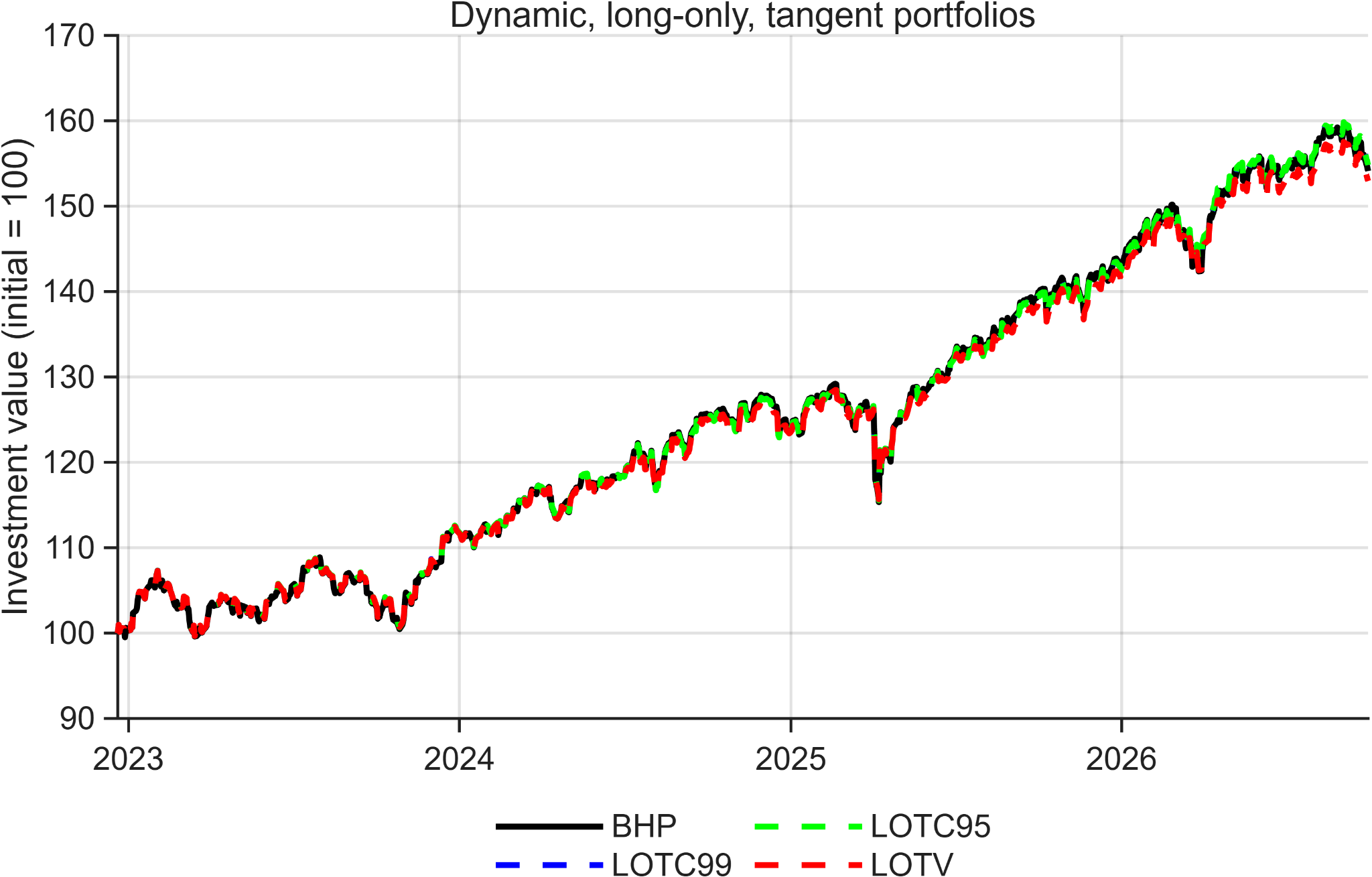}
\vspace{0.3em}
\includegraphics[width=0.47\textwidth]{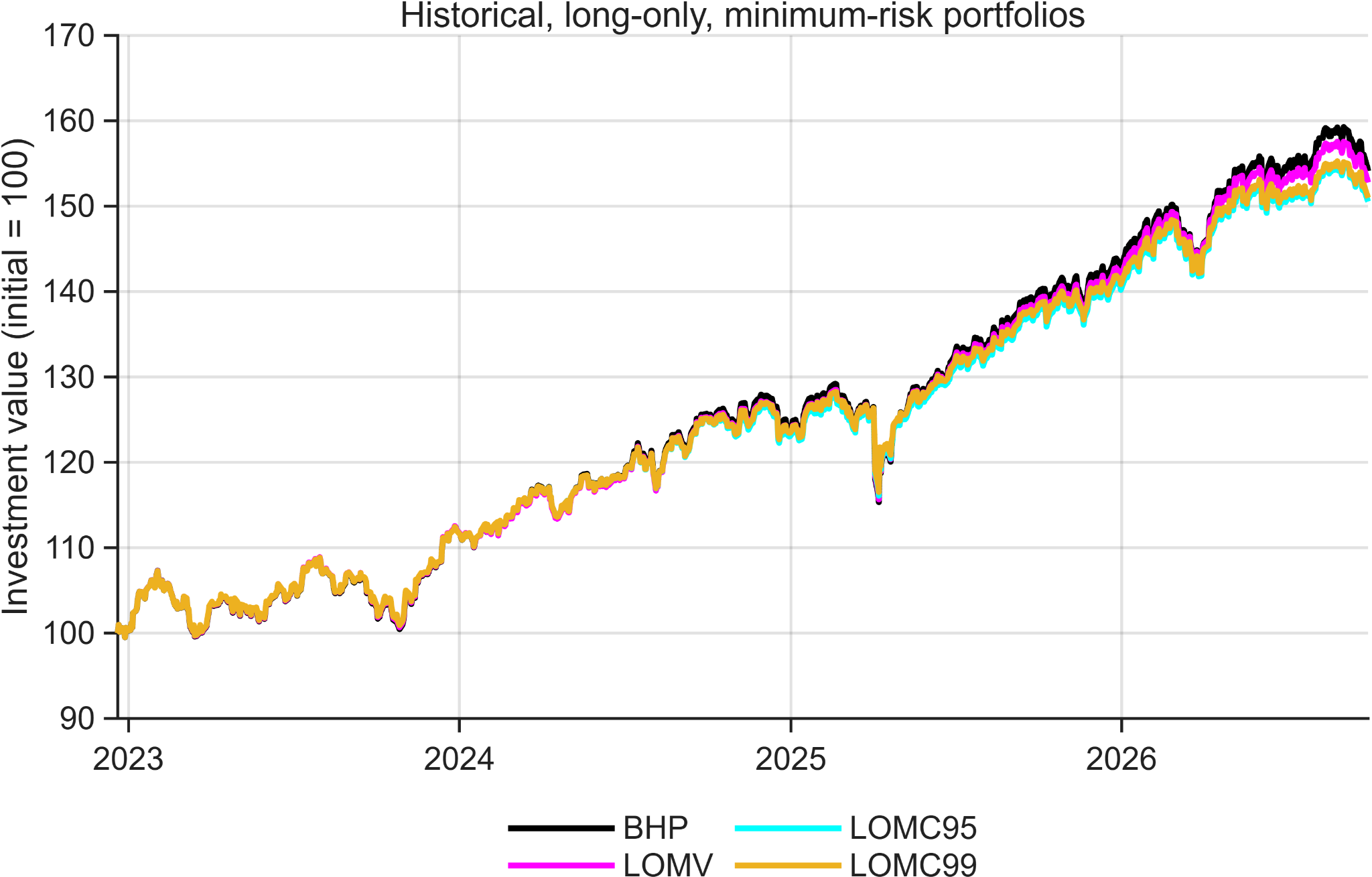}\hfill
\includegraphics[width=0.47\textwidth]{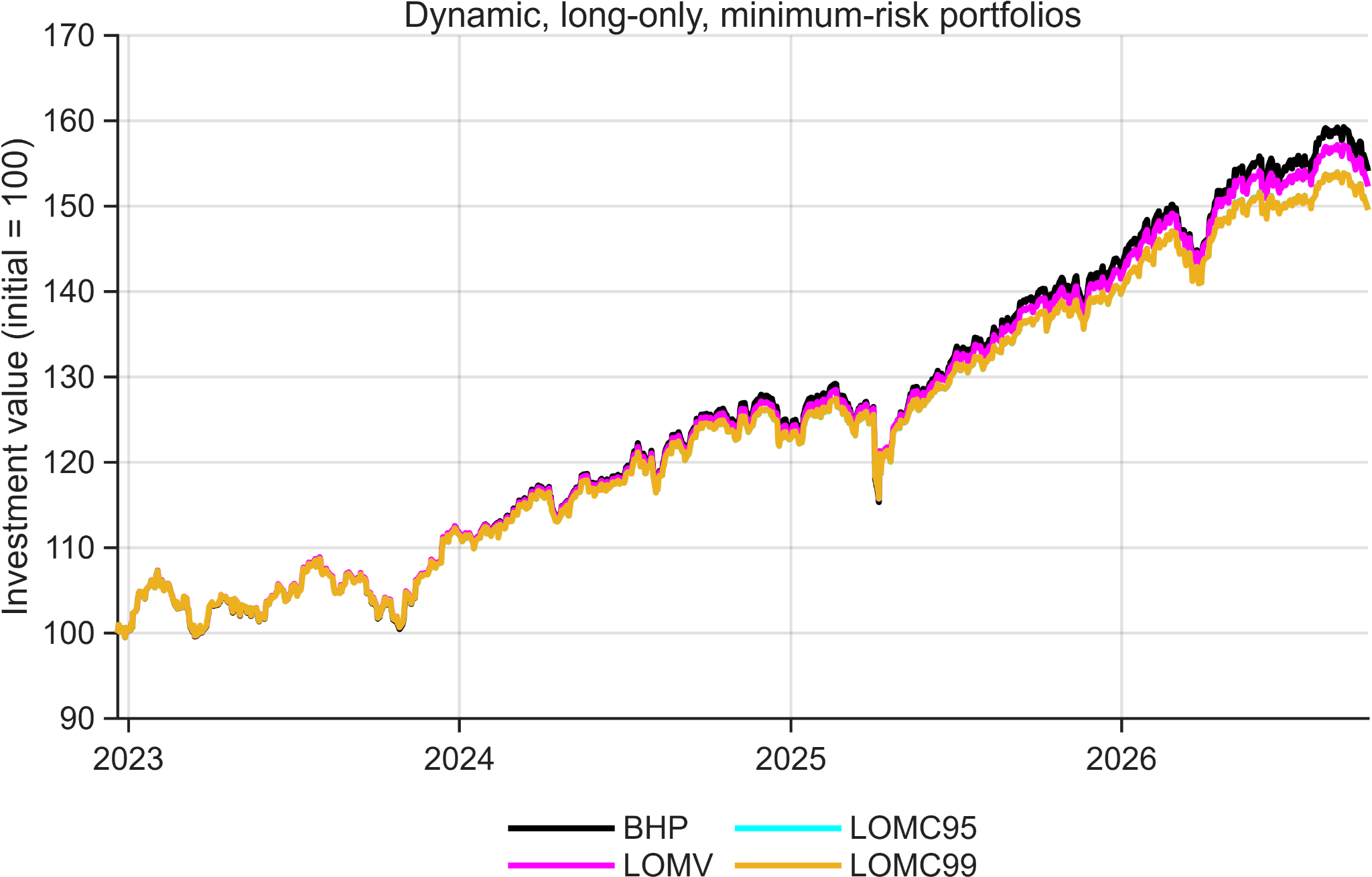}
\vspace{0.3em}
\includegraphics[width=0.47\textwidth]{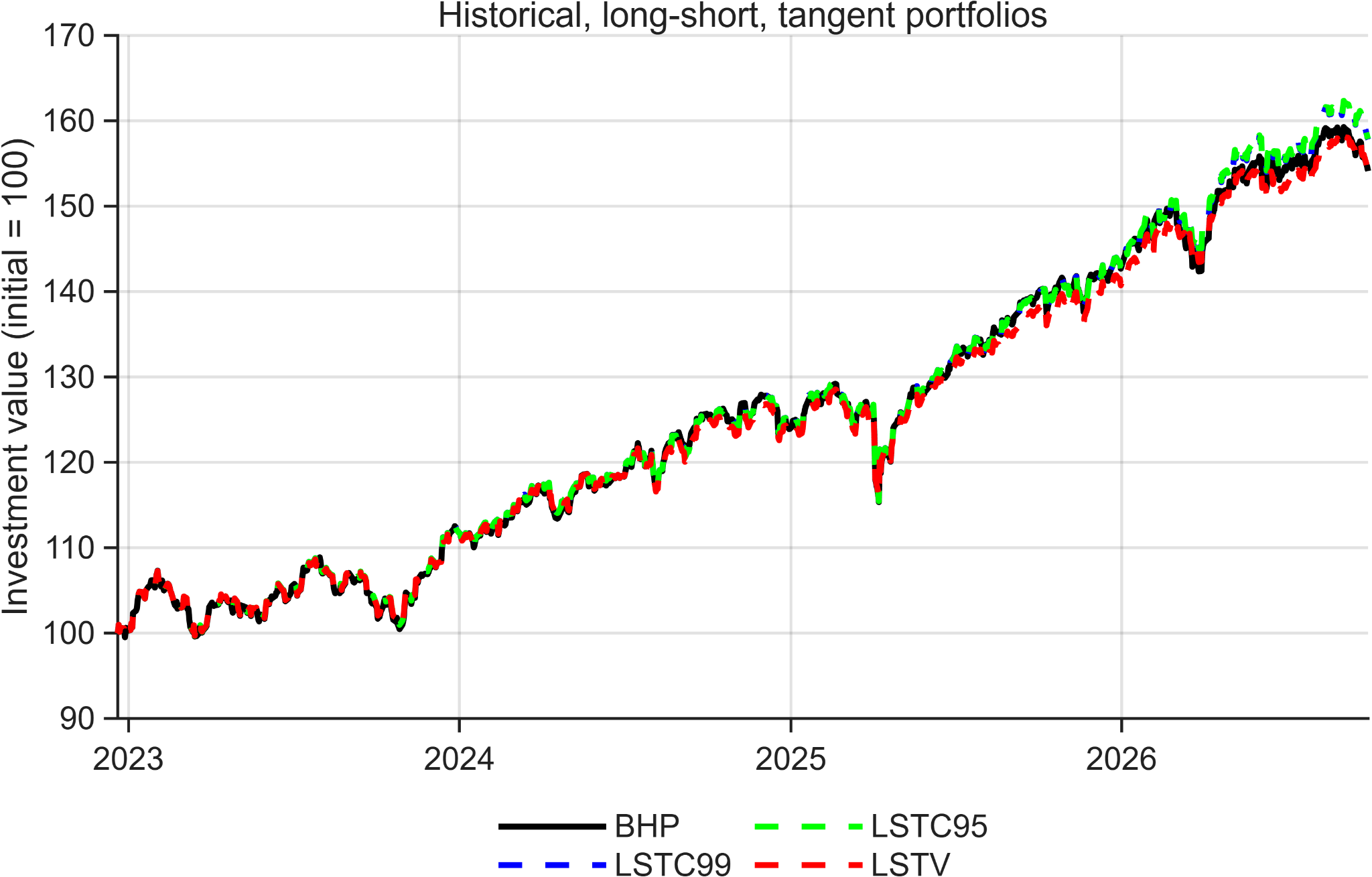}\hfill
\includegraphics[width=0.47\textwidth]{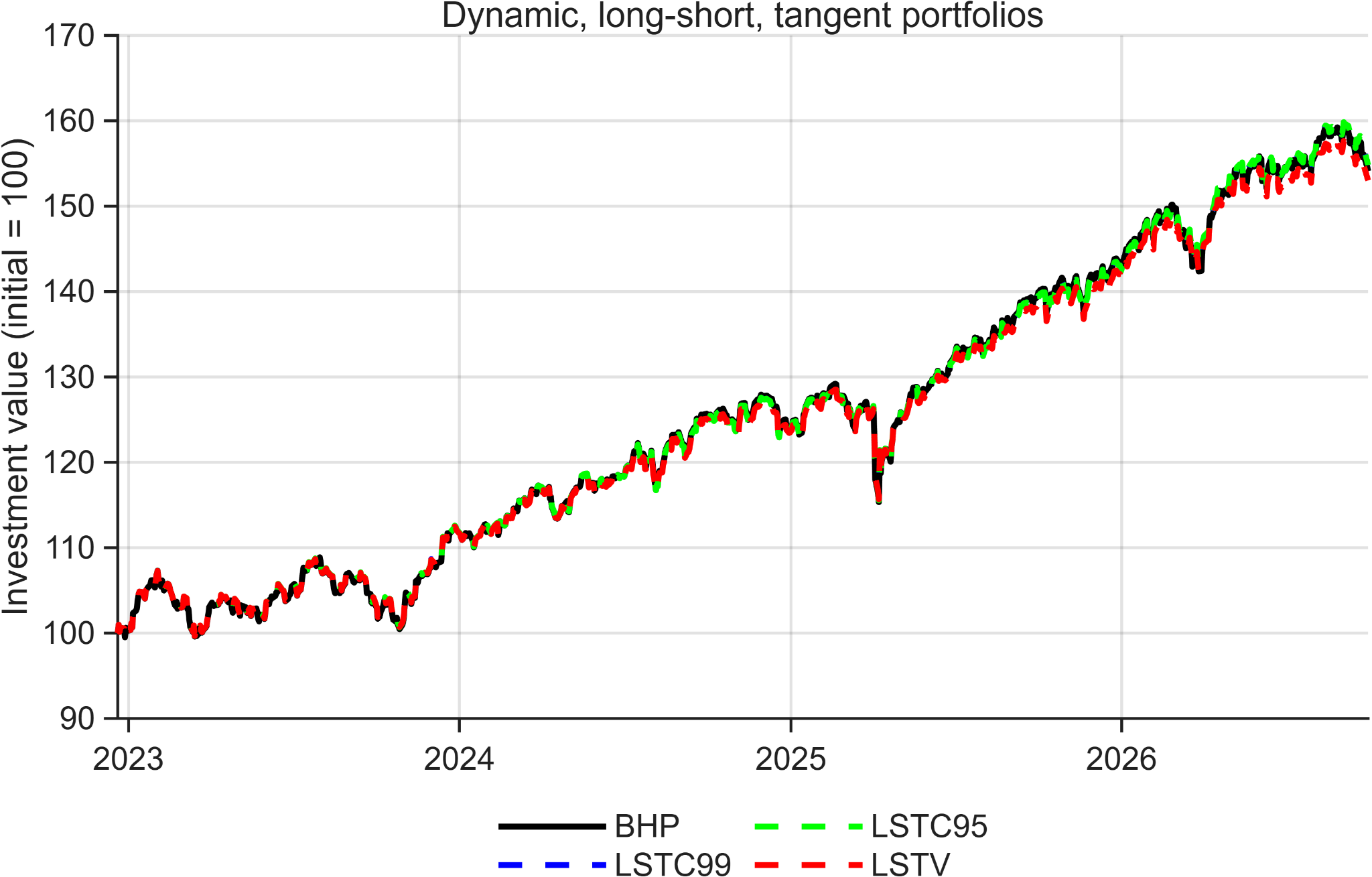}
\vspace{0.3em}
\includegraphics[width=0.47\textwidth]{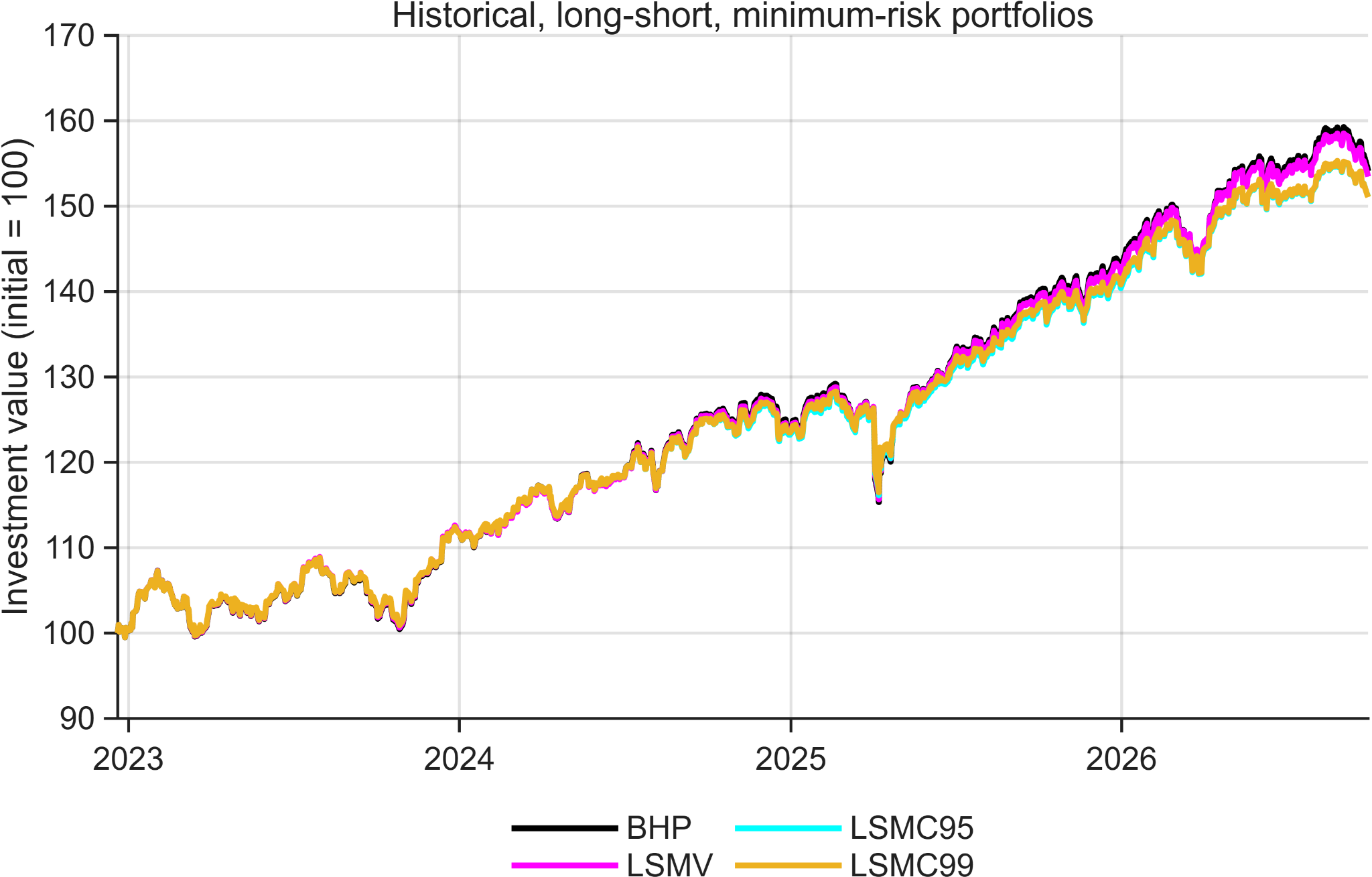}\hfill
\includegraphics[width=0.47\textwidth]{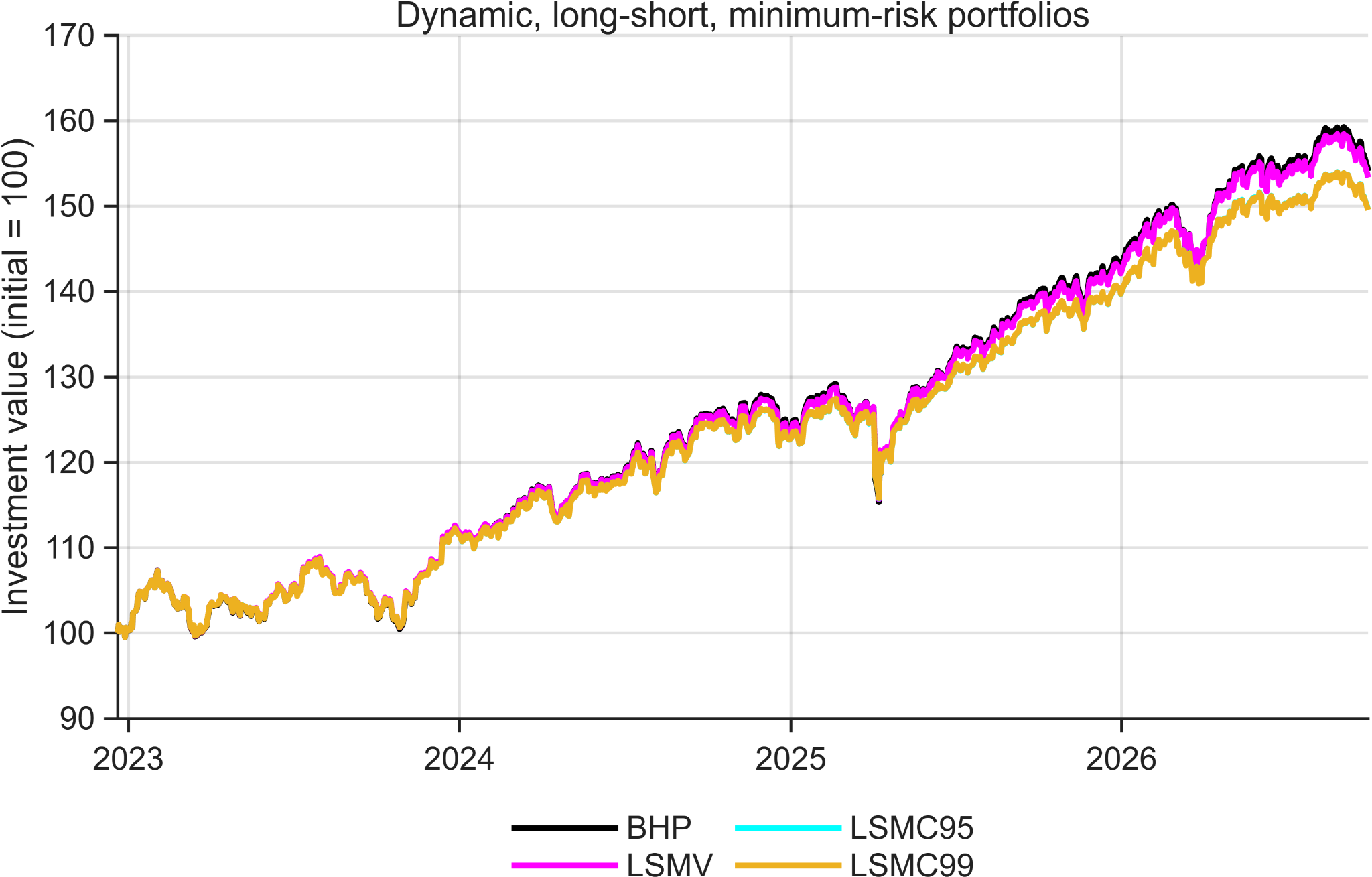}
\caption{Baseline gross portfolio-value paths (initial investment USD~100), 21~December~2022--30~September~2026. Historical results appear on the left and dynamic results on the right. Rows show long--only tangent, long--only minimum-risk, long--short tangent, and long--short minimum-risk strategies. BHP denotes the common initially equal-weighted unrebalanced benchmark. Portfolio weights selected using preceding estimation observations are applied to subsequent realized ETF returns.}
\label{fig:portfolio_value_compare}
\end{figure}

For portfolio outcomes, Figure~\ref{fig:portfolio_value_compare} displays cumulative realized gross investment values of the 24 baseline strategies and the initially equal-weighted buy-and-hold benchmark. Historical and dynamic results are displayed in matched columns, and long--only and long--short strategies are grouped consistently. These are gross, next-observation return-attribution paths, not net-of-cost executable trading records.

The historical and conditional wealth paths were often similar for the same objective and investment bounds (Figure~\ref{fig:portfolio_value_compare}). Both methods earned returns on the same underlying ETF market movements, and the tight adjustment limit left relatively little room for their target allocations to diverge quickly. Differences among portfolio objectives were at times more visible than differences between the two return-input methods. These gross wealth paths are interpreted together with turnover and modeled costs below.

\subsection{Full-sample reward--risk performance}\label{sec:performance}
For each of the 24 optimized portfolios, we evaluated four realized daily reward--risk ratios over the common 946-session sample. Let $x_{p,t}=r_{p,t}-r_{f,t}$ be the excess arithmetic return, with the daily risk-free return computed from the most recent available three-month Treasury yield as $(1+y_t/100)^{1/252}-1$. We report the Sharpe ratio $\overline{x}/s(x)$ \citep{Sharpe1994}, the positive-part quadratic downside ratio (SS$_2$), $\overline{\max(x,0)}/[\overline{\max(-x,0)^2}]^{1/2}$, the Rachev ratio at 95/95, $\mathrm{TM}_{0.05}(x)/\mathrm{TM}_{0.05}(-x)$, and the STAR ratio at 95\%, $\overline{x}/\mathrm{TM}_{0.05}(-x)$; $\mathrm{TM}_{p}$ denotes the exact upper-tail average including a fractional cutoff observation. These are unannualized daily ratios and are calculated identically in the full-sample performance audit and paired bootstrap.

\begin{table}[htbp]\centering
\scriptsize
\caption{Realized full-sample reward--risk ratios for the 24 baseline portfolios (946 sessions).}\label{tab:rr2026}
\setlength{\tabcolsep}{3pt}
\begin{tabular}{lllrrrr}\toprule

Method & Bounds & Portfolio & Sharpe & SS$_2$ & Rachev & STAR \\
\midrule

Hist. & LO & MV & 0.05216 & 0.5518 & 0.9355 & 0.02335 \\

Dynamic & LO & MV & 0.05179 & 0.5509 & 0.9344 & 0.02314 \\

Hist. & LO & TV & 0.05478 & 0.5521 & 0.9309 & 0.02432 \\

Dynamic & LO & TV & 0.05289 & 0.5517 & 0.9376 & 0.02364 \\

Hist. & LO & MC95 & 0.05184 & 0.5536 & 0.9408 & 0.02310 \\

Dynamic & LO & MC95 & 0.05056 & 0.5512 & 0.9373 & 0.02248 \\

Hist. & LO & TC95 & 0.05658 & 0.5526 & 0.9313 & 0.02526 \\

Dynamic & LO & TC95 & 0.05367 & 0.5520 & 0.9357 & 0.02406 \\

Hist. & LO & MC99 & 0.05226 & 0.5537 & 0.9391 & 0.02329 \\

Dynamic & LO & MC99 & 0.05049 & 0.5511 & 0.9376 & 0.02244 \\

Hist. & LO & TC99 & 0.05648 & 0.5526 & 0.9323 & 0.02524 \\

Dynamic & LO & TC99 & 0.05366 & 0.5520 & 0.9358 & 0.02405 \\

Hist. & LS & MV & 0.05226 & 0.5518 & 0.9368 & 0.02346 \\

Dynamic & LS & MV & 0.05219 & 0.5517 & 0.9365 & 0.02342 \\

Hist. & LS & TV & 0.05547 & 0.5527 & 0.9418 & 0.02478 \\

Dynamic & LS & TV & 0.05290 & 0.5517 & 0.9381 & 0.02366 \\

Hist. & LS & MC95 & 0.05242 & 0.5540 & 0.9415 & 0.02338 \\

Dynamic & LS & MC95 & 0.05056 & 0.5512 & 0.9373 & 0.02248 \\

Hist. & LS & TC95 & 0.05690 & 0.5535 & 0.9367 & 0.02544 \\

Dynamic & LS & TC95 & 0.05367 & 0.5520 & 0.9357 & 0.02406 \\

Hist. & LS & MC99 & 0.05228 & 0.5535 & 0.9411 & 0.02332 \\

Dynamic & LS & MC99 & 0.05049 & 0.5511 & 0.9376 & 0.02244 \\

Hist. & LS & TC99 & 0.05678 & 0.5535 & 0.9379 & 0.02541 \\

Dynamic & LS & TC99 & 0.05366 & 0.5520 & 0.9358 & 0.02405 \\

\bottomrule\end{tabular}\end{table}
\FloatBarrier

Table~\ref{tab:rr2026} shows higher observed reward--risk ratios for historical inputs in most matched comparisons, particularly for several tangent portfolios. A tangent objective trades estimated risk against expected excess return, while a minimum-risk objective gives greater weight to limiting dispersion or tail loss. As a result, the portfolio with the lowest estimated risk need not provide the highest realized return per unit of risk. The rankings in Table~\ref{tab:rr2026} describe this sample; the paired tests in Section~\ref{sec:bootstrap2026} assess how far they support broader inference.

\subsection{Descriptive moments and realized return densities}\label{sec:moments}
\begin{figure}[htbp]\centering
\includegraphics[width=0.96\textwidth]{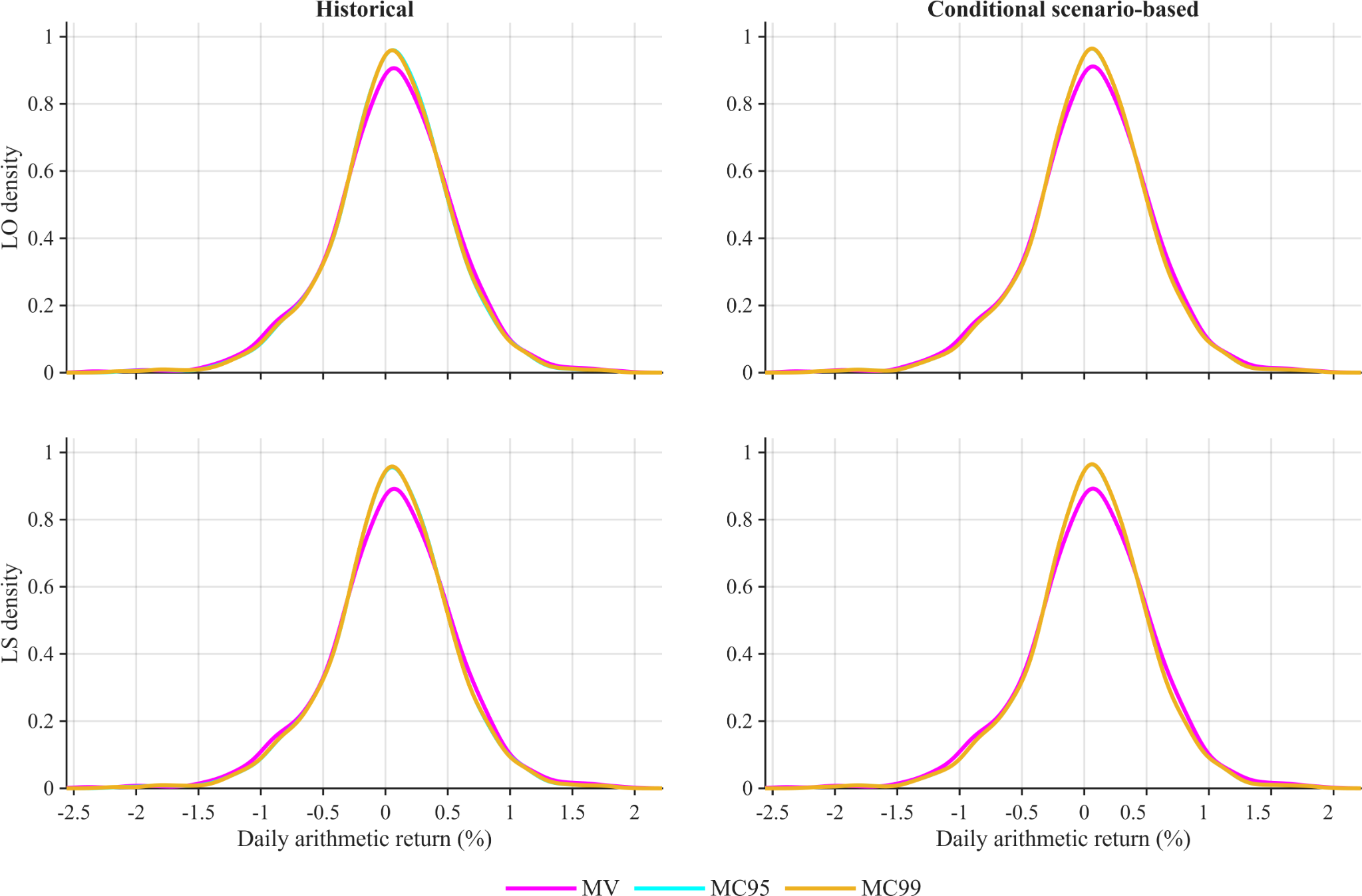}
\caption{Kernel density estimates for daily gross returns of MV, MC95 and MC99 portfolios. Long--only results appear above limited long--short results; historical estimates appear to the left of dynamic estimates. Bandwidth is selected separately for each realized series.}\label{fig:kdemin2026}\end{figure}

\begin{figure}[htbp]\centering
\includegraphics[width=0.96\textwidth]{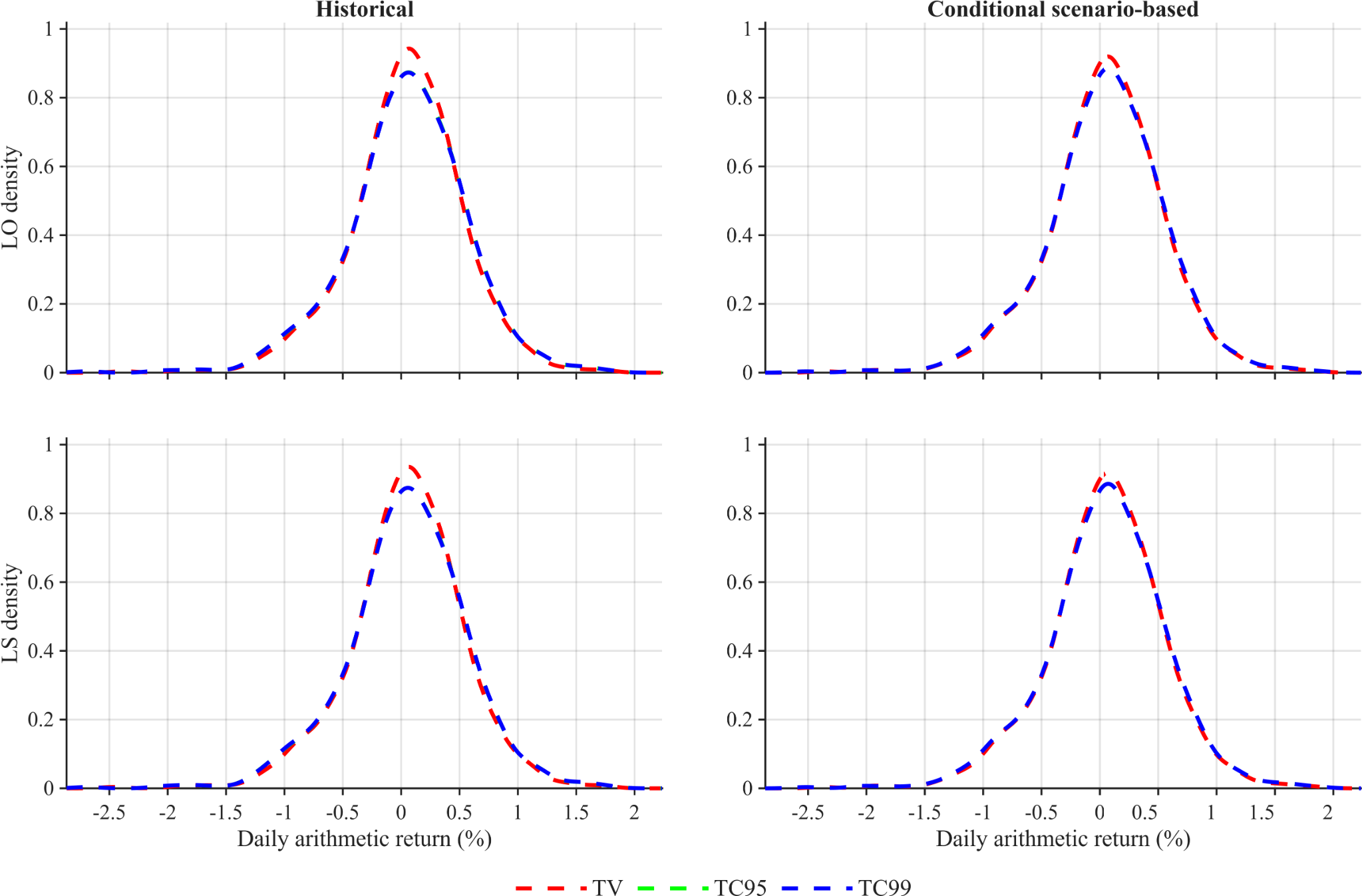}
\caption{Kernel density estimates for realized daily gross returns of TV, TC95 and TC99 portfolios, with matched scales across method and investment-constraint panels.}\label{fig:kdetan2026}\end{figure}

Tables~\ref{tab:moments2026} and \ref{tab:momentsls2026} described the central tendency, dispersion, and tail shape of the realized gross returns. Means and standard deviations are reported in units of $10^{-4}$, skewness in units of $10^{-2}$, and excess kurtosis without scaling. The moments were calculated from the same realized daily portfolio-return histories used for performance evaluation, not from simulated scenarios. Figures~\ref{fig:kdemin2026} and \ref{fig:kdetan2026} complement these summaries with empirical return densities.

\begin{table}[htbp]\centering
\tiny
\caption{Descriptive moments of daily realized gross portfolio returns: long--only (H = historical; D = dynamic).}\label{tab:moments2026}
\setlength{\tabcolsep}{3pt}
\begin{tabular}{lrrrrrrrr}\toprule

Strategy & H LO mean & sd & skew & ex.kurt. & D LO mean & sd & skew & ex.kurt. \\
\midrule

MC95 & 4.45 & 51.03 & 3.24 & 9.66 & 4.39 & 51.03 & 2.82 & 9.59 \\

MV & 4.63 & 54.00 & 3.20 & 10.71 & 4.59 & 53.74 & 2.52 & 10.69 \\

MC99 & 4.49 & 51.21 & 1.13 & 9.47 & 4.39 & 51.01 & 2.53 & 9.55 \\

TC95 & 4.97 & 55.85 & 3.72 & 12.53 & 4.77 & 55.22 & 4.01 & 11.46 \\

TC99 & 4.97 & 55.94 & 4.01 & 12.56 & 4.77 & 55.22 & 4.04 & 11.46 \\

TV & 4.66 & 52.10 & 2.15 & 11.02 & 4.63 & 53.41 & 4.06 & 10.94 \\

\bottomrule\end{tabular}\end{table}
\FloatBarrier

\begin{table}[htbp]\centering
\tiny
\caption{Descriptive moments of daily realized gross portfolio returns: limited long--short. Same units as Table~\ref{tab:moments2026}.}\label{tab:momentsls2026}
\setlength{\tabcolsep}{3pt}
\begin{tabular}{lrrrrrrrr}\toprule

Strategy & H LS mean & sd & skew & ex.kurt. & D LS mean & sd & skew & ex.kurt. \\
\midrule

MC95 & 4.50 & 51.25 & 4.25 & 9.76 & 4.39 & 51.03 & 2.82 & 9.59 \\

MV & 4.68 & 54.90 & 3.37 & 10.87 & 4.67 & 54.86 & 3.34 & 10.89 \\

MC99 & 4.49 & 51.32 & 1.47 & 9.60 & 4.39 & 51.01 & 2.53 & 9.55 \\

TC95 & 4.98 & 55.77 & 4.40 & 12.43 & 4.77 & 55.22 & 4.01 & 11.46 \\

TC99 & 4.98 & 55.82 & 4.65 & 12.41 & 4.77 & 55.22 & 4.04 & 11.46 \\

TV & 4.73 & 52.56 & 4.23 & 11.53 & 4.64 & 53.50 & 4.14 & 10.97 \\

\bottomrule\end{tabular}\end{table}
\FloatBarrier

\FloatBarrier

\FloatBarrier

The realized return densities in Figures~\ref{fig:kdemin2026} and \ref{fig:kdetan2026} overlapped for many historical--conditional pairs. The shared market returns and limited movement in portfolio weights provide plausible reasons for this similarity. Tables~\ref{tab:moments2026} and \ref{tab:momentsls2026} also report skewness and excess kurtosis, indicating that average returns and standard deviations do not fully describe the realized return distributions. Similar fitted densities should not be interpreted as evidence that the input distributions or their predictive accuracy were equivalent.

\subsection{Moving-window reward--risk and downside diagnostics}\label{sec:moving2026}
\begin{figure}[htbp]\centering
\includegraphics[width=0.96\textwidth]{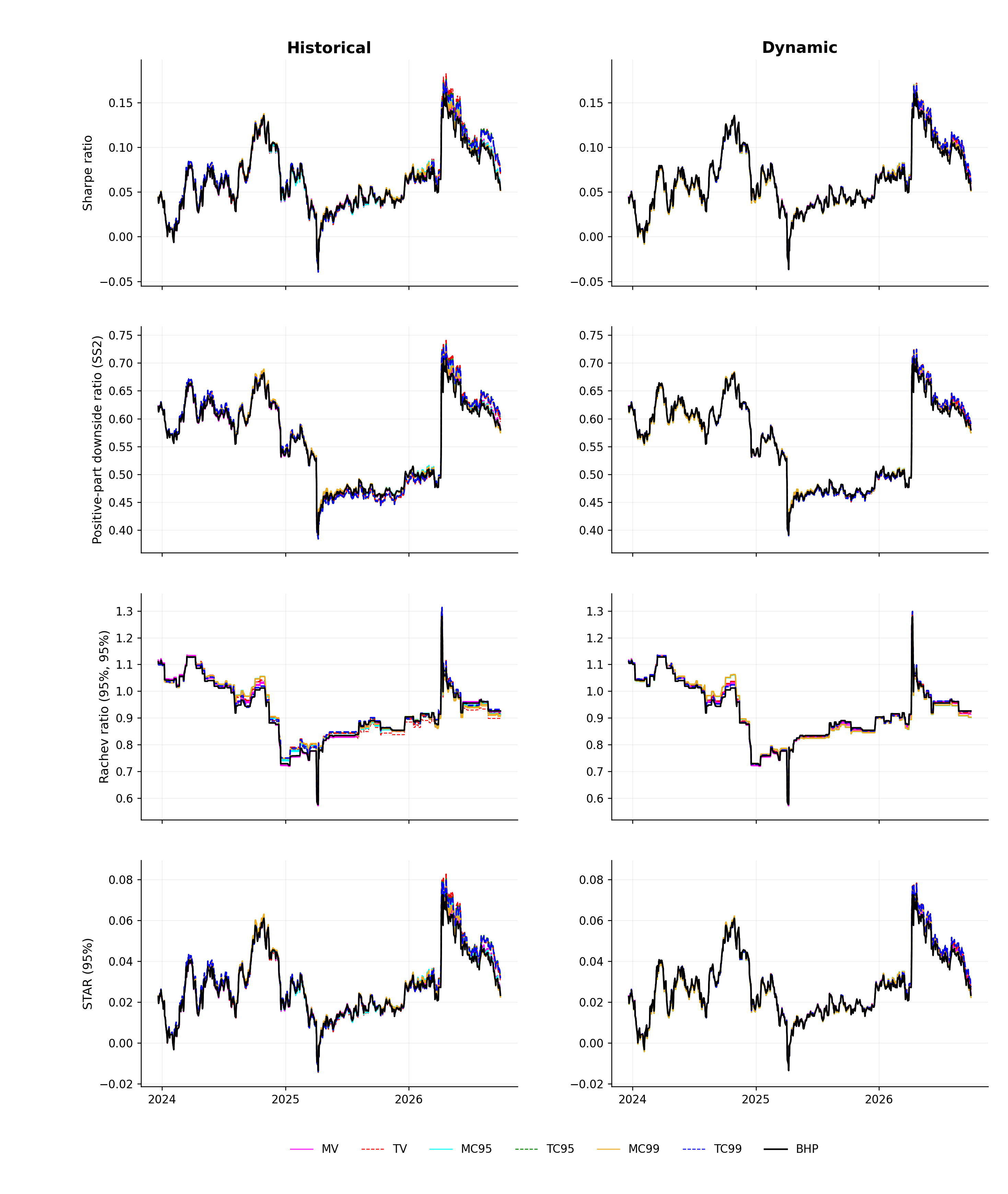}
\caption{252-observation moving realized reward--risk ratios. Historical portfolios (left) and dynamic portfolios (right); colors indicate objectives, line thickness distinguishes LO and LS, and BHP denotes the buy-and-hold benchmark. Both method panels use identical axes within each metric.}\label{fig:movingrr2026}\end{figure}

\begin{figure}[htbp]\centering
\includegraphics[width=0.96\textwidth]{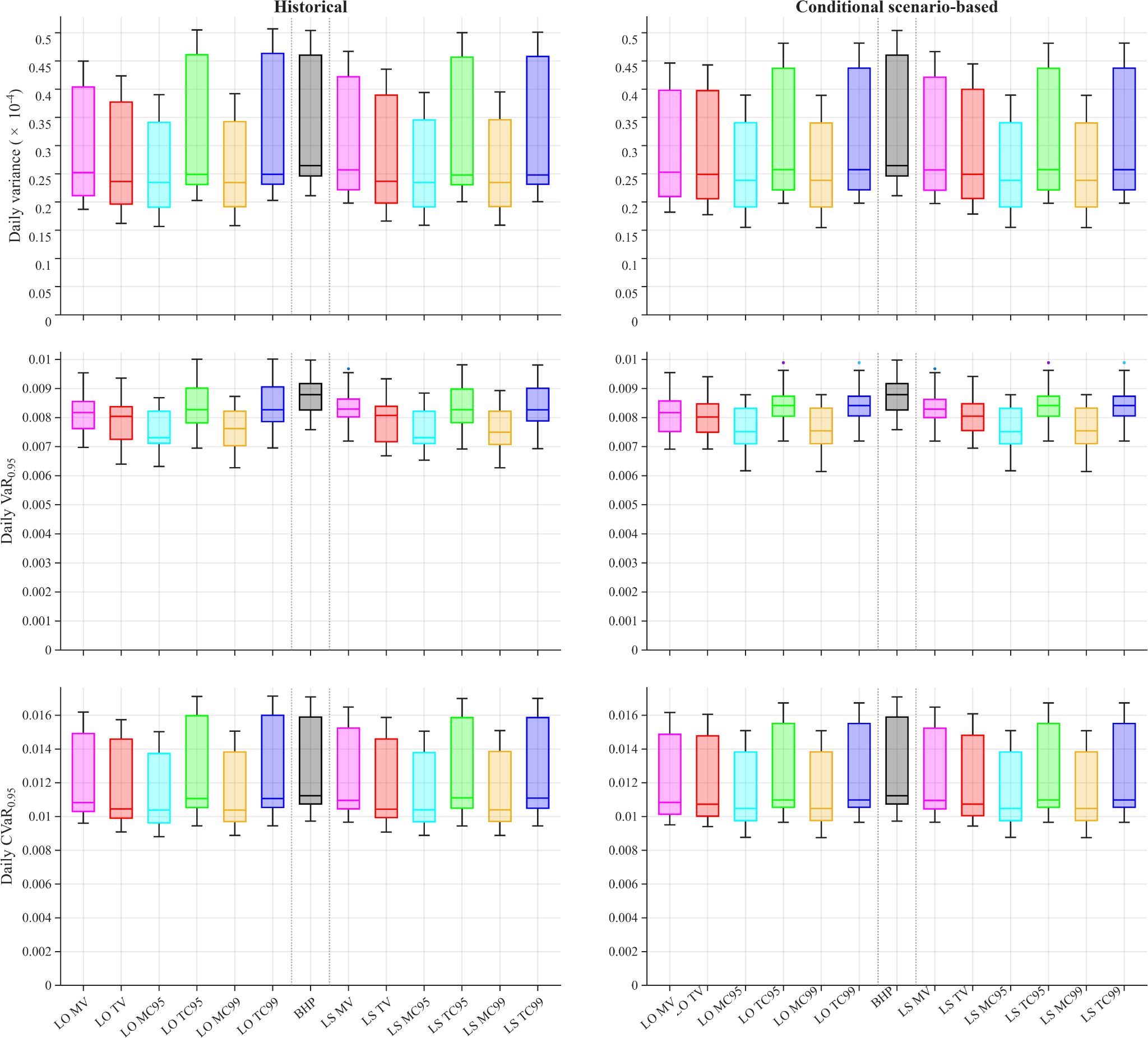}
\caption{Distributions over 695 moving windows of realized daily variance ($10^{-4}$ units), 95\% loss VaR, and 95\% loss CVaR. Method panels have matched scales for each diagnostic; boxplots describe variability over overlapping moving windows and are not independent samples.}\label{fig:movingrisk2026}\end{figure}

A full-sample measure can conceal changes in performance across market conditions. Figure~\ref{fig:movingrr2026} follows reward--risk ratios over 252-session moving windows, while Figure~\ref{fig:movingrisk2026} summarizes moving variance and realized downside-loss measures. The 946-session evaluation yielded 695 overlapping windows. Reward--risk measures used daily excess returns, whereas variance, loss value at risk (VaR), and loss CVaR used gross returns. Loss VaR was the empirical 95th percentile of negative returns, and loss CVaR was the corresponding average upper-tail loss with a fractional cutoff. These overlapping windows describe changes through time rather than independent observations.

\FloatBarrier

\FloatBarrier

\begin{table}[htbp]\centering
\scriptsize
\caption{Medians across 695 overlapping moving diagnostic windows for minimum-risk portfolios; daily variance displayed in $10^{-4}$ units, CVaR in decimal daily-return units.}\label{tab:riskmed2026}
\setlength{\tabcolsep}{3pt}
\begin{tabular}{llrrrr}\toprule

Constraint & Strategy & Historical var. & Dynamic var. & Historical CVaR & Dynamic CVaR \\
\midrule

LO & MV & 0.2522 & 0.2529 & 0.01083 & 0.01084 \\

LO & MC95 & 0.2347 & 0.2385 & 0.01038 & 0.01048 \\

LO & MC99 & 0.2346 & 0.2385 & 0.01038 & 0.01048 \\

LS & MV & 0.2569 & 0.2568 & 0.01096 & 0.01095 \\

LS & MC95 & 0.2347 & 0.2385 & 0.01040 & 0.01048 \\

LS & MC99 & 0.2347 & 0.2385 & 0.01040 & 0.01048 \\

\bottomrule\end{tabular}\end{table}
\FloatBarrier

The moving-window risk measures did not show a consistent reduction in realized losses from using conditional scenarios. In Table~\ref{tab:riskmed2026}, historical minimum-CVaR portfolios generally had slightly lower median realized variance and CVaR than their conditional counterparts, while the minimum-variance comparisons were mixed. An optimizer minimizes risk estimated from its inputs, but subsequent losses reflect the returns that actually occurred. The evidence therefore does not show that the conditional approach systematically improved downside protection over the examined period.

\subsection{Portfolio turnover, exposures, and transaction costs}\label{sec:turnnet2026}
A stable sequence of optimized weights need not describe a low-trading strategy. Table~\ref{tab:turn2026} distinguishes one-way changes between successive target weights from one-way trades needed to restore target weights after asset prices have changed. The first measure was restricted by the baseline optimization cap; the second was not. These turnover summaries excluded the first evaluated allocation because its preceding holdings were not available. Annualized figures in Table~\ref{tab:turn2026} were calculated from the mean of the 945 observed transitions multiplied by 252. Short exposure measures negative ETF weights in absolute value, while gross exposure sums absolute weights. Table~\ref{tab:grossnet2026} then relates the implementation measure to illustrative trading-cost adjustments.

\begin{table}[htbp]\centering
\scriptsize
\caption{Baseline annualized target-adjustment and drift-adjusted implementation turnover and average short exposure; percentages are indicated in the headings.}\label{tab:turn2026}
\setlength{\tabcolsep}{3pt}
\begin{tabular}{lllrrrr}\toprule

Method & Bounds & Strategy & Target TO (\%) & Impl. TO (\%) & Short exposure (\%) & Cap breaches \\
\midrule

Hist. & LO & MV & 1.47 & 41.48 & 0.00 & 0 \\

Hist. & LO & TV & 3.24 & 43.60 & 0.00 & 0 \\

Hist. & LO & MC95 & 3.33 & 41.51 & 0.00 & 0 \\

Hist. & LO & TC95 & 3.33 & 44.75 & 0.00 & 0 \\

Hist. & LO & MC99 & 3.33 & 41.67 & 0.00 & 0 \\

Hist. & LO & TC99 & 3.33 & 44.79 & 0.00 & 0 \\

Dyn. & LO & MV & 1.71 & 41.48 & 0.00 & 0 \\

Dyn. & LO & TV & 3.13 & 42.28 & 0.00 & 0 \\

Dyn. & LO & MC95 & 3.33 & 41.49 & 0.00 & 0 \\

Dyn. & LO & TC95 & 3.33 & 43.41 & 0.00 & 0 \\

Dyn. & LO & MC99 & 3.33 & 41.53 & 0.00 & 0 \\

Dyn. & LO & TC99 & 3.33 & 43.42 & 0.00 & 0 \\

Hist. & LS & MV & 0.09 & 41.72 & 0.00 & 0 \\

Hist. & LS & TV & 3.16 & 46.07 & 1.94 & 0 \\

Hist. & LS & MC95 & 3.33 & 42.73 & 0.91 & 0 \\

Hist. & LS & TC95 & 3.33 & 46.25 & 1.48 & 0 \\

Hist. & LS & MC99 & 3.33 & 42.57 & 0.70 & 0 \\

Hist. & LS & TC99 & 3.33 & 46.15 & 1.45 & 0 \\

Dyn. & LS & MV & 0.12 & 41.71 & 0.00 & 0 \\

Dyn. & LS & TV & 3.09 & 42.22 & 0.00 & 0 \\

Dyn. & LS & MC95 & 3.33 & 41.49 & 0.00 & 0 \\

Dyn. & LS & TC95 & 3.33 & 43.41 & 0.00 & 0 \\

Dyn. & LS & MC99 & 3.33 & 41.53 & 0.00 & 0 \\

Dyn. & LS & TC99 & 3.33 & 43.42 & 0.00 & 0 \\

\bottomrule\end{tabular}\end{table}
\FloatBarrier

Table~\ref{tab:turn2026} shows that relatively small changes in optimized target weights could coexist with much larger implementation turnover. ETF price movements altered the proportions held between rebalancing decisions, requiring trades to restore the targets even when the optimizer changed its desired allocation very little. The baseline limit constrained the first quantity, not the second. For a manager assessing trading needs and costs, target-adjustment turnover alone would therefore provide an incomplete measure of portfolio activity.

Transaction costs were assessed at ten basis points per unit of one-way implementation turnover ($c=0.001$). For the allocation selected using information through date $t$, the cost-adjusted return attributed to the following observed session was
\begin{equation}
r^{\mathrm{net}}_{p,t+1}
=\left(1-c\kappa_t\right)
\left(1+\boldsymbol w_t^{\mathsf T}\boldsymbol r_{t+1}\right)-1.
\label{eq:cost_adjusted_next_return}
\end{equation}
Here $\kappa_t$ is defined in Equation~\eqref{eq:implementation_turnover}, and the second factor is the gross return in Equation~\eqref{eq:gross_next_return}. The cost charge was thus associated with the target earning the next-session return. In the saved daily files, the turnover and return for that holding period share the same next-session date label, rather than the decision-date index used in these equations. Initial formation costs, borrowing and financing expenses on short positions, and execution frictions beyond the assumed linear cost were omitted. The buy-and-hold benchmark was not rebalanced and had no modeled post-formation turnover costs. For the gross/net summary, implementation turnover was annualized over the 946 evaluated returns with day-one turnover set to zero; Table~\ref{tab:turn2026} instead annualized the mean over the 945 observed transitions. Both denominators are stated to distinguish the two summaries.

\begin{table}[htbp]\centering
\scriptsize
\caption{Gross and transaction-cost-adjusted performance (946 observations). Geometric returns and volatility are annualized; daily loss CVaR95 and maximum drawdown are expressed in percent.}\label{tab:grossnet2026}
\setlength{\tabcolsep}{3pt}
\begin{tabular}{lllrrrrr}\toprule

Method & Bounds & Portfolio & Gross return (\%) & Net return (\%) & Gross vol. (\%) & Net CVaR95 (\%) & MDD net (\%) \\
\midrule

Hist. & LO & MV & 11.95 & 11.90 & 8.57 & 1.188 & 9.99 \\

Hist. & LO & TV & 12.09 & 12.04 & 8.27 & 1.155 & 9.86 \\

Hist. & LO & MC95 & 11.51 & 11.47 & 8.10 & 1.126 & 9.14 \\

Hist. & LO & TC95 & 12.90 & 12.85 & 8.87 & 1.233 & 10.99 \\

Hist. & LO & MC99 & 11.60 & 11.55 & 8.13 & 1.131 & 9.15 \\

Hist. & LO & TC99 & 12.89 & 12.84 & 8.88 & 1.234 & 11.01 \\

Hist. & LS & MV & 12.09 & 12.04 & 8.71 & 1.205 & 10.21 \\

Hist. & LS & TV & 12.25 & 12.20 & 8.34 & 1.158 & 10.03 \\

Hist. & LS & MC95 & 11.62 & 11.58 & 8.14 & 1.130 & 9.21 \\

Hist. & LS & TC95 & 12.93 & 12.88 & 8.85 & 1.229 & 10.93 \\

Hist. & LS & MC99 & 11.61 & 11.57 & 8.15 & 1.132 & 9.20 \\

Hist. & LS & TC99 & 12.92 & 12.87 & 8.86 & 1.229 & 10.94 \\

Dyn. & LO & MV & 11.86 & 11.81 & 8.53 & 1.184 & 9.94 \\

Dyn. & LO & TV & 11.98 & 11.94 & 8.48 & 1.177 & 9.98 \\

Dyn. & LO & MC95 & 11.33 & 11.28 & 8.10 & 1.130 & 9.16 \\

Dyn. & LO & TC95 & 12.35 & 12.30 & 8.77 & 1.214 & 10.49 \\

Dyn. & LO & MC99 & 11.32 & 11.27 & 8.10 & 1.129 & 9.15 \\

Dyn. & LO & TC99 & 12.35 & 12.30 & 8.77 & 1.214 & 10.49 \\

Dyn. & LS & MV & 12.07 & 12.02 & 8.71 & 1.204 & 10.21 \\

Dyn. & LS & TV & 12.00 & 11.95 & 8.49 & 1.178 & 10.01 \\

Dyn. & LS & MC95 & 11.33 & 11.28 & 8.10 & 1.130 & 9.16 \\

Dyn. & LS & TC95 & 12.35 & 12.30 & 8.77 & 1.214 & 10.49 \\

Dyn. & LS & MC99 & 11.32 & 11.27 & 8.10 & 1.129 & 9.15 \\

Dyn. & LS & TC99 & 12.35 & 12.30 & 8.77 & 1.214 & 10.49 \\

BHP & -- & -- & 12.21 & 12.21 & 9.01 & 1.246 & 10.72 \\

\bottomrule\end{tabular}\end{table}
\FloatBarrier

The cost-adjusted returns in Table~\ref{tab:grossnet2026} remained close to the gross results under the assumed linear trading-cost schedule. This follows partly from the modest cost rate applied to the estimated implementation turnover and should not be interpreted as evidence that daily rebalancing is inexpensive in practice. Bid--ask spreads, market impact, financing, and short-borrowing charges were outside the calculation. The unrebalanced buy-and-hold benchmark incurred no modeled post-formation trading charges but did not maintain the optimized strategies' exposures.

\subsection{Paired bootstrap inference and block-length sensitivity}\label{sec:bootstrap2026}
For each of the 12 matched constraint--objective portfolios and four reward--risk measures, we calculated $\Delta=\mathrm{Ratio}_{\mathrm{dynamic}}-\mathrm{Ratio}_{\mathrm{historical}}$, yielding 48 paired comparisons. The stationary bootstrap \citep{PolitisRomano1994} resampled common dated indices jointly for the portfolios and their aligned risk-free returns. We used 5,000 replicates for each expected block length of 10, 20, and 40 sessions (seed 12345), and calculated percentile confidence intervals and centered-bootstrap two-sided $p$-values. Holm family-wise error and Benjamini--Hochberg false-discovery-rate adjustments \citep{Holm1979,BenjaminiHochberg1995} were applied separately across the 48 tests at each block length. Appendix~\ref{app:bootstrap_pvalues2026} reports individual raw and adjusted $p$-values; Appendix Tables~\ref{tab:bootstrap_alt_10_LO_2026}--\ref{tab:bootstrap_alt_40_LS_2026} give the complete results for alternative block lengths.

\begin{table}[htbp]
\centering
\caption{Dynamic-minus-historical daily reward--risk differences and unadjusted 95\% paired stationary-bootstrap percentile confidence intervals. Results use 5,000 resamples with expected block length 20; the interval for each comparison is shown in parentheses. Adjusted statistical conclusions are reported separately.}
\label{tab:paired_bootstrap_detailed_2026}
\setlength{\tabcolsep}{3pt}
\renewcommand{\arraystretch}{1.10}
\scriptsize
\begin{tabular}{lrrrr}
\toprule
Portfolio & $\Delta$Sharpe & $\Delta$SS$_2$ & $\Delta$Rachev$_{95/95}$ & $\Delta$STAR$_{95}$ \\
\midrule
LO MV & \shortstack{$-0.00037$\\{\tiny $(-0.00100, 0.00025)$}} & \shortstack{$-0.00082$\\{\tiny $(-0.00182, 0.00029)$}} & \shortstack{$-0.00113$\\{\tiny $(-0.00390, 0.00224)$}} & \shortstack{$-0.00021$\\{\tiny $(-0.00050, 0.00010)$}} \\
LO TV & \shortstack{$-0.00189$\\{\tiny $(-0.00737, 0.00271)$}} & \shortstack{$-0.00041$\\{\tiny $(-0.00774, 0.00520)$}} & \shortstack{$0.00667$\\{\tiny $(-0.00697, 0.02034)$}} & \shortstack{$-0.00068$\\{\tiny $(-0.00312, 0.00140)$}} \\
LO MC95 & \shortstack{$-0.00127$\\{\tiny $(-0.00304, 0.00041)$}} & \shortstack{$-0.00239$\\{\tiny $(-0.00486, -0.00036)$}} & \shortstack{$-0.00354$\\{\tiny $(-0.01005, 0.00301)$}} & \shortstack{$-0.00063$\\{\tiny $(-0.00147, 0.00015)$}} \\
LO TC95 & \shortstack{$-0.00291$\\{\tiny $(-0.00816, 0.00121)$}} & \shortstack{$-0.00061$\\{\tiny $(-0.01018, 0.00549)$}} & \shortstack{$0.00442$\\{\tiny $(-0.01159, 0.02027)$}} & \shortstack{$-0.00120$\\{\tiny $(-0.00363, 0.00073)$}} \\
LO MC99 & \shortstack{$-0.00178$\\{\tiny $(-0.00360, 0.00002)$}} & \shortstack{$-0.00266$\\{\tiny $(-0.00536, -0.00005)$}} & \shortstack{$-0.00153$\\{\tiny $(-0.00924, 0.00820)$}} & \shortstack{$-0.00084$\\{\tiny $(-0.00169, 0.00002)$}} \\
LO TC99 & \shortstack{$-0.00282$\\{\tiny $(-0.00795, 0.00123)$}} & \shortstack{$-0.00057$\\{\tiny $(-0.01015, 0.00542)$}} & \shortstack{$0.00353$\\{\tiny $(-0.01241, 0.01928)$}} & \shortstack{$-0.00119$\\{\tiny $(-0.00360, 0.00072)$}} \\
LS MV & \shortstack{$-0.00008$\\{\tiny $(-0.00014, -0.00001)$}} & \shortstack{$-0.00016$\\{\tiny $(-0.00028, -0.00003)$}} & \shortstack{$-0.00030$\\{\tiny $(-0.00063, 0.00012)$}} & \shortstack{$-0.00004$\\{\tiny $(-0.00007, -4.63\times10^{-6})$}} \\
LS TV & \shortstack{$-0.00257$\\{\tiny $(-0.00839, 0.00219)$}} & \shortstack{$-0.00099$\\{\tiny $(-0.00989, 0.00494)$}} & \shortstack{$-0.00370$\\{\tiny $(-0.01751, 0.01047)$}} & \shortstack{$-0.00112$\\{\tiny $(-0.00381, 0.00110)$}} \\
LS MC95 & \shortstack{$-0.00185$\\{\tiny $(-0.00417, 0.00026)$}} & \shortstack{$-0.00285$\\{\tiny $(-0.00599, -0.00012)$}} & \shortstack{$-0.00421$\\{\tiny $(-0.01228, 0.00434)$}} & \shortstack{$-0.00090$\\{\tiny $(-0.00200, 0.00009)$}} \\
LS TC95 & \shortstack{$-0.00323$\\{\tiny $(-0.00912, 0.00148)$}} & \shortstack{$-0.00146$\\{\tiny $(-0.01189, 0.00528)$}} & \shortstack{$-0.00096$\\{\tiny $(-0.01969, 0.01665)$}} & \shortstack{$-0.00138$\\{\tiny $(-0.00424, 0.00086)$}} \\
LS MC99 & \shortstack{$-0.00179$\\{\tiny $(-0.00405, 0.00028)$}} & \shortstack{$-0.00246$\\{\tiny $(-0.00596, 0.00037)$}} & \shortstack{$-0.00354$\\{\tiny $(-0.01282, 0.00663)$}} & \shortstack{$-0.00087$\\{\tiny $(-0.00193, 0.00011)$}} \\
LS TC99 & \shortstack{$-0.00311$\\{\tiny $(-0.00884, 0.00145)$}} & \shortstack{$-0.00151$\\{\tiny $(-0.01168, 0.00502)$}} & \shortstack{$-0.00205$\\{\tiny $(-0.02086, 0.01557)$}} & \shortstack{$-0.00136$\\{\tiny $(-0.00413, 0.00084)$}} \\
\bottomrule
\end{tabular}
\end{table}
\FloatBarrier

Most dynamic-minus-historical differences in Table~\ref{tab:paired_bootstrap_detailed_2026} were negative, consistent with the observed reward--risk ordering. Some individual unadjusted percentile intervals excluded zero. However, no paired comparison remained significant after the stated Holm or Benjamini--Hochberg adjustments across the full family of tests (Appendix Table~\ref{tab:bootstrap2026}). The estimated differences therefore do not support a claim of general historical superiority, nor does the absence of adjusted rejections establish equivalence. Individual adjusted $p$-values and alternative block-length results are reported in Appendices~\ref{app:bootstrap_pvalues2026} and \ref{app:bootstrap_alternative_blocks}.

\subsection{Tail behavior and Pareto approximation}\label{sec:tails2026}
The tail analysis compared the observed decay of positive loss and gain magnitudes for the optimized portfolios and the buy-and-hold benchmark. Log-survival slopes and Hill estimates were calculated using upper-tail thresholds of 5\%, 7\%, and 10\% of each portfolio's 946 out-of-sample returns \citep{deHaanFerreira2006,Resnick2007}. At the 7\% threshold, 67 observations entered each fit. Confidence intervals for the OLS tail-index estimates used 400 stationary-bootstrap resamples with an expected block length of 20. These estimates describe selected empirical tails; they do not establish an exact Pareto distribution or calibrated future tail-loss probabilities. Table~\ref{tab:pareto2026} and Figure~\ref{fig:paretotc992026} present representative results, with another illustration in Appendix Figure~\ref{fig:paretomc992026}.

\begin{table}[htbp]\centering
\scriptsize
\caption{Illustrative 7\% upper-tail OLS and Hill estimates for selected portfolios and the buy-and-hold benchmark. The 95\% stationary-bootstrap confidence intervals apply to the OLS tail-index estimates, not the Hill estimates.}\label{tab:pareto2026}
\setlength{\tabcolsep}{3pt}
\begin{tabular}{llrrl}\toprule

Portfolio & Tail & OLS index & Hill index & 95\% OLS CI \\
\midrule

Historical LS TC99 & Loss & 2.81 & 2.91 & [2.07, 4.70] \\

Historical LS TC99 & Gain & 2.86 & 3.34 & [2.08, 4.56] \\

Dynamic LS TC99 & Loss & 2.89 & 3.03 & [2.07, 4.67] \\

Dynamic LS TC99 & Gain & 2.87 & 3.29 & [2.05, 4.76] \\

Historical LS MC99 & Loss & 2.94 & 2.96 & [2.15, 4.58] \\

Historical LS MC99 & Gain & 3.01 & 3.63 & [2.18, 4.88] \\

Dynamic LS MC99 & Loss & 2.95 & 2.93 & [2.20, 4.60] \\

Dynamic LS MC99 & Gain & 2.97 & 3.51 & [2.11, 4.72] \\

BHP & Loss & 2.90 & 3.04 & [2.09, 4.79] \\

BHP & Gain & 2.87 & 3.43 & [2.12, 4.59] \\

\bottomrule\end{tabular}\end{table}
\FloatBarrier

The illustrative tail estimates in Table~\ref{tab:pareto2026} were broadly similar for the matched portfolios, and the OLS confidence intervals were wide relative to those differences. The plots describe the observed upper tails at selected thresholds; they do not establish a common Pareto law or reliable predictive tail calibration. We therefore place more weight on the realized loss comparisons when discussing the risk-management implications of the study.

\begin{figure}[htbp]\centering
\includegraphics[width=0.96\textwidth]{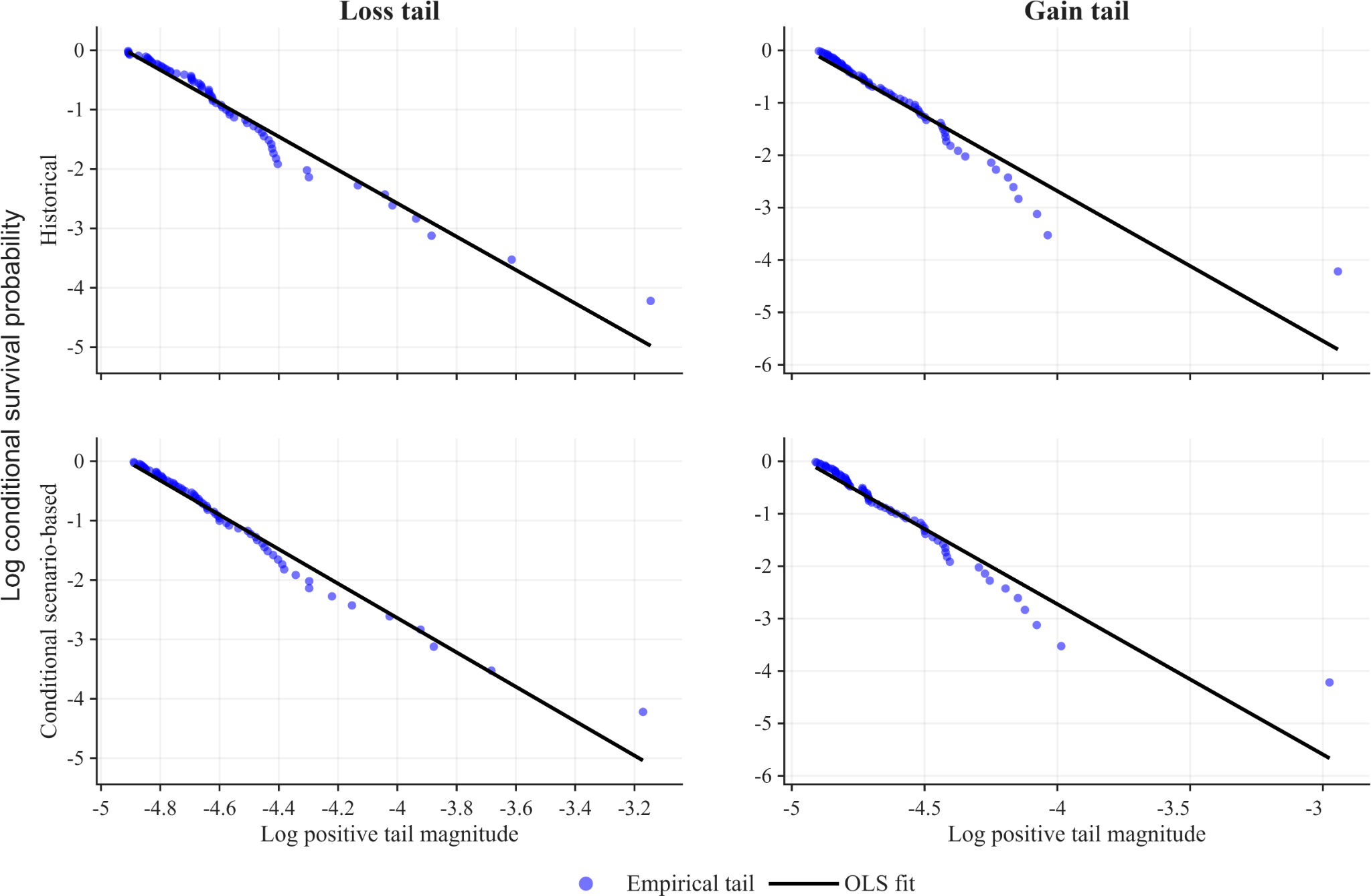}
\caption{Empirical log-survival points and OLS fitted lines for positive loss and gain tails of the LS TC99 portfolio, historical above and dynamic below. Fits use the largest 67 positive observations in each respective tail.}\label{fig:paretotc992026}\end{figure}
\FloatBarrier

The allocation and sensitivity results show how the portfolio rules affected holdings and trading, providing context for the similar realized outcomes.

\subsection{Allocation and implementation diagnostics}\label{sec:allocation_figs2026}
\begin{figure}[htbp]\centering
\includegraphics[width=\textwidth]{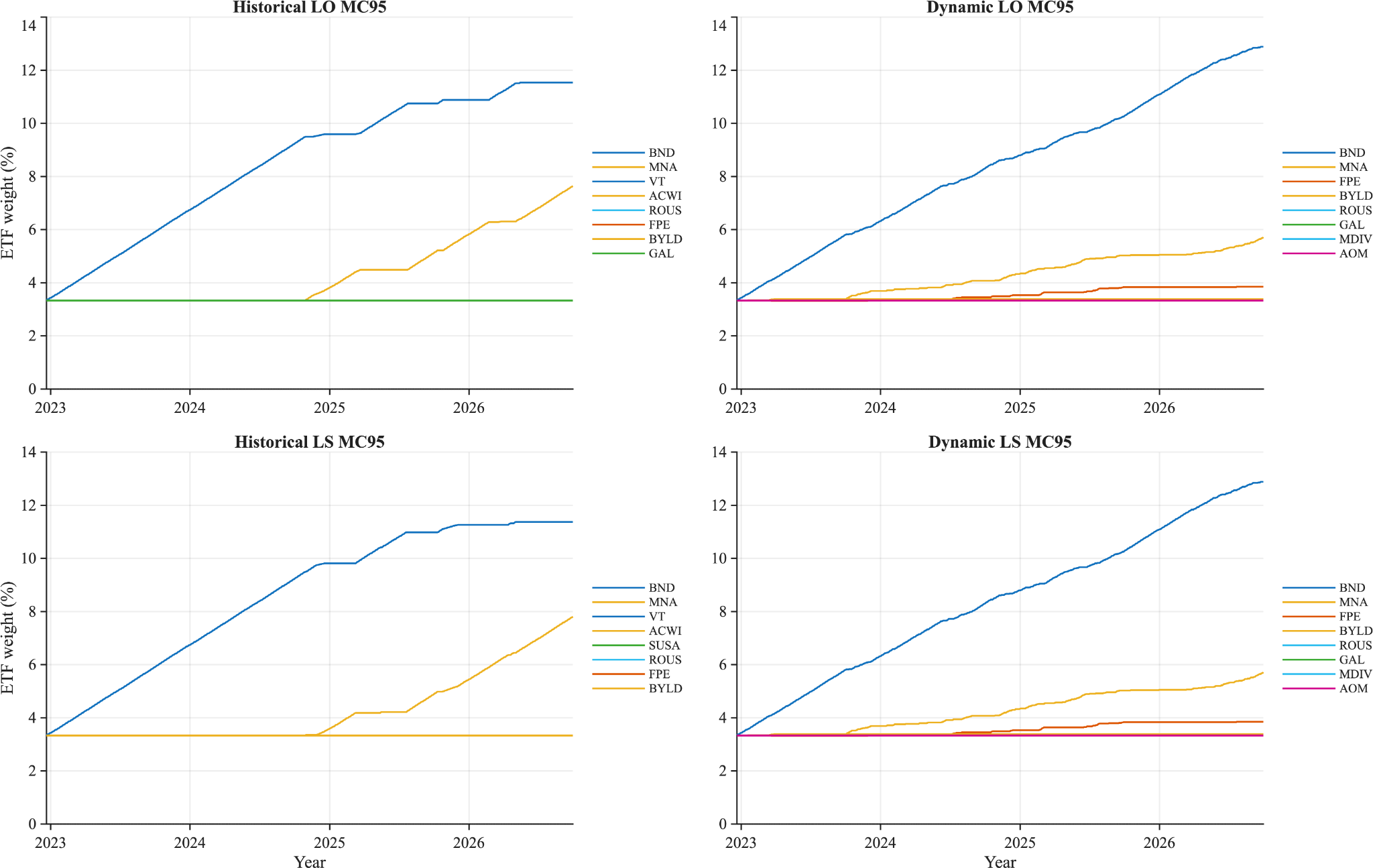}
\caption{Evolution of selected ETF portfolio weights in the baseline MC95 portfolios. Historical (left) and dynamic (right) are compared under long-only (top) and limited long-short (bottom) constraints. The eight ETFs with the largest mean absolute weight are shown separately for each portfolio; overlapping series remain distinct assets. Weights are percentages of portfolio net asset value.}\label{fig:alloc_mc95_2026}
\end{figure}

\begin{figure}[htbp]\centering
\includegraphics[width=\textwidth]{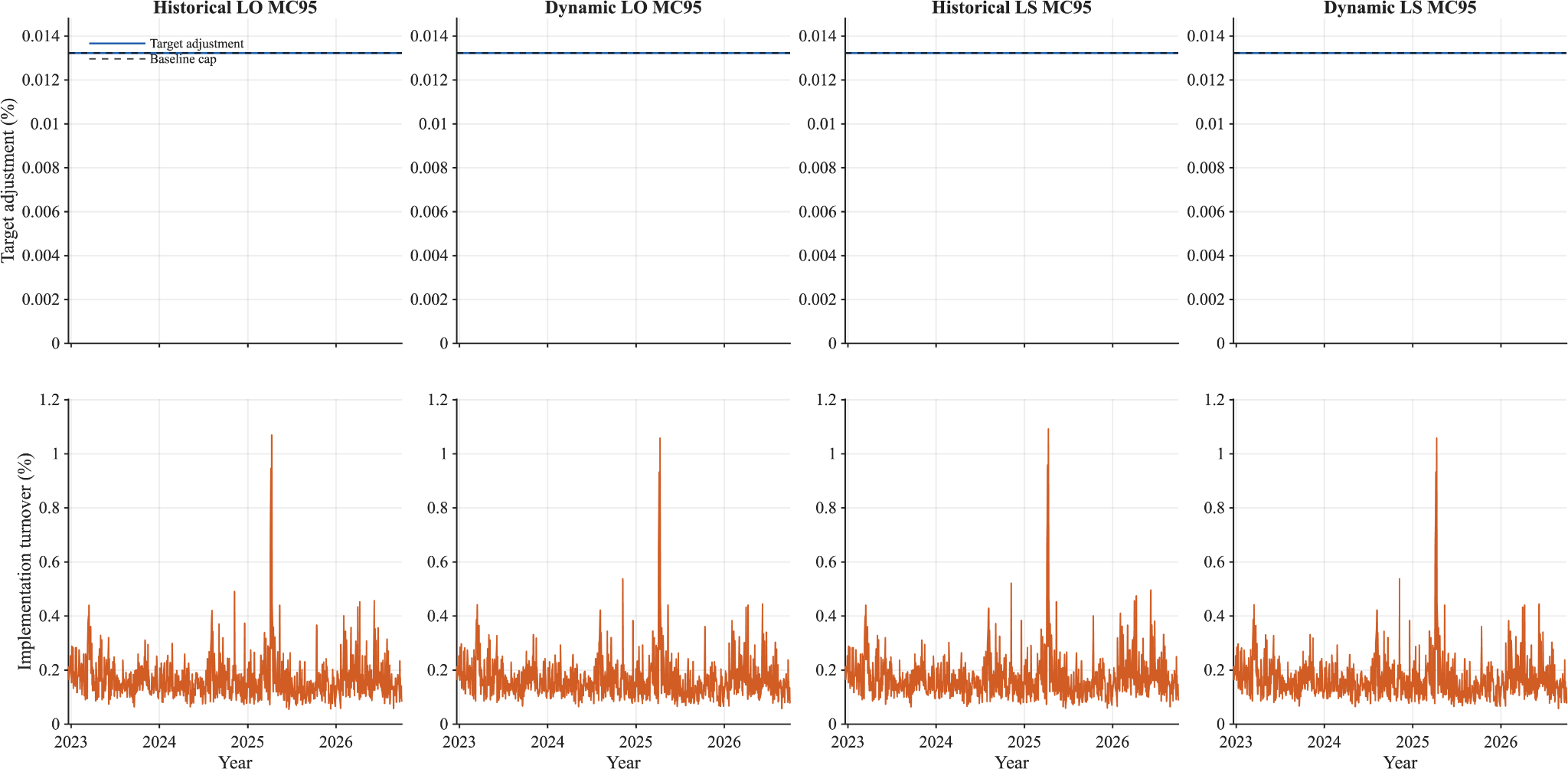}
\caption{Baseline MC95 target adjustments (top) and drift-adjusted implementation turnover (bottom) for historical and dynamic LO and LS portfolios, as percentages of net asset value per trading day. The dashed line is the baseline cap on target adjustments; implementation turnover is not subject to this cap. The two top-row curves nearly coincide because adjustments frequently approach the cap. Initial portfolio formation is excluded.}\label{fig:turnover_paths_mc95_2026}
\end{figure}

\begin{figure}[htbp]\centering
\includegraphics[width=\textwidth]{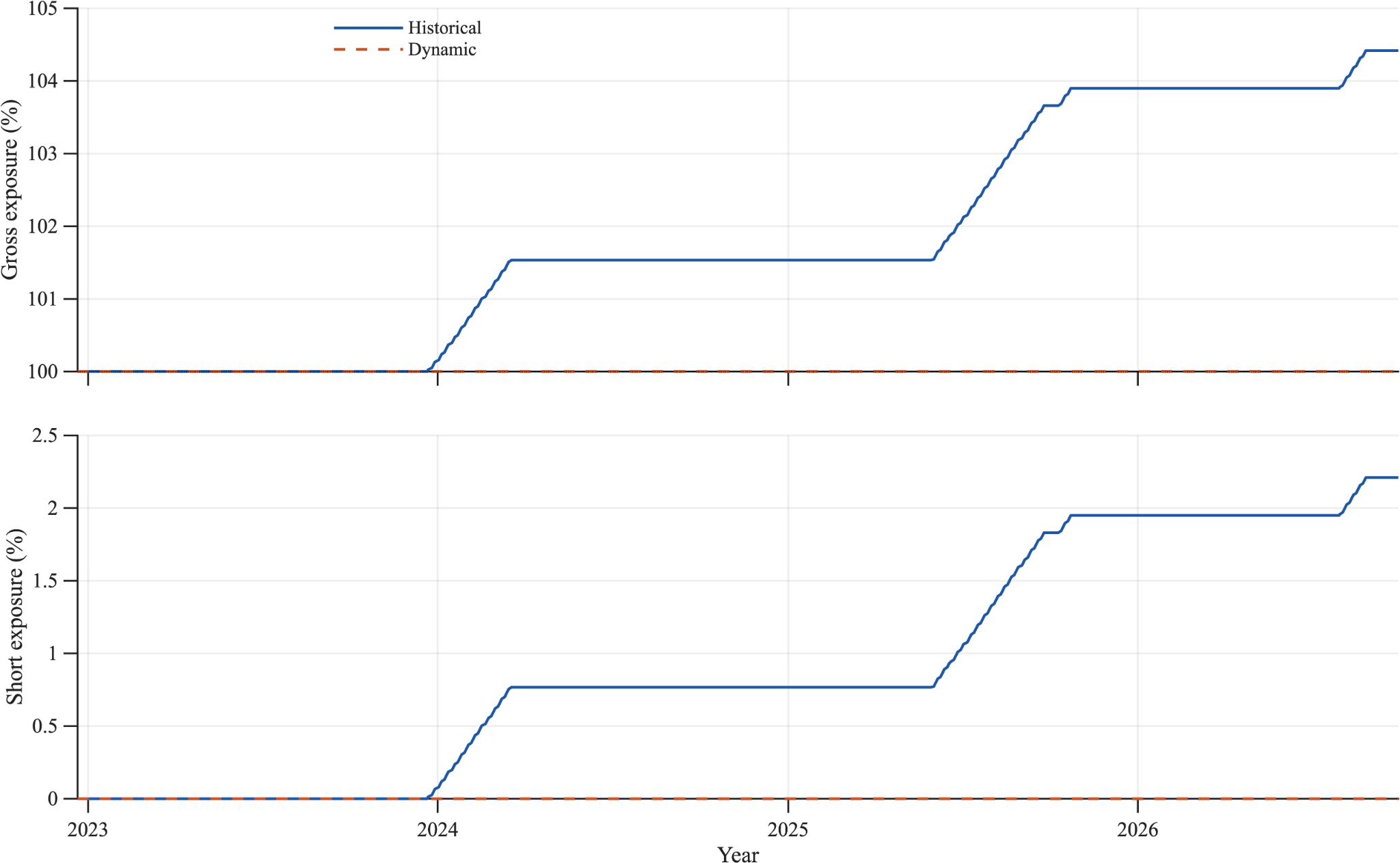}
\caption{Gross and short exposure for baseline historical and dynamic long-short MC95 portfolios. Gross exposure equals the sum of absolute ETF weights; short exposure equals the sum of absolute negative ETF weights, each expressed as a percentage of net asset value.}\label{fig:short_mc95_2026}
\end{figure}

Figure~\ref{fig:alloc_mc95_2026} follows the principal ETF holdings in the minimum-CVaR 95\% strategies. The target-adjustment restriction allowed positions to move only gradually from their common equal-weight start. The resulting allocations nevertheless differed for selected ETFs, suggesting that the return inputs influenced portfolio choices even when the rule limited the speed of adjustment.
Figure~\ref{fig:turnover_paths_mc95_2026} separates changes in targets from the trades needed to restore them after the underlying ETFs earned different returns. The rebalancing series varied even when target adjustments remained close to their permitted limit. Thus, a rule that restrains optimization-driven changes does not necessarily impose a comparable limit on the volume of trades required to maintain the portfolio.
Figure~\ref{fig:short_mc95_2026} compares gross and short exposures for the limited long--short MC95 portfolios. Although both input methods permitted negative positions, the historical portfolio developed some short exposure while the conditional portfolio remained effectively long-only. Permitting shorts therefore did not guarantee that the optimizer would use them within the narrow daily adjustment allowance.
\FloatBarrier

\subsection{Portfolio allocations and concentration}\label{sec:allocation_reconciled_2026}
Table~\ref{tab:allocation_reconciled_mc95_2026} relates the MC95 return and risk findings to the ETF positions actually held. It reports the direct holdings in BND and BYLD, the largest ETF weight, concentration, and short and gross exposure. These are holdings in traded ETFs, not estimates of the underlying asset classes within mixed-allocation funds. Appendix Table~\ref{tab:allocation_reconciled_all_2026} provides the corresponding measures for all baseline strategies.

\begin{table}[htbp]\centering\scriptsize
\caption{MC95 baseline ETF allocations and concentration (946 dates). BND, BYLD, combined bond, largest ETF, gross exposure and short exposure are percentages of portfolio capital; HHI is a unitless concentration index.}\label{tab:allocation_reconciled_mc95_2026}
\resizebox{\textwidth}{!}{%
\begin{tabular}{lllrrrrrrr}\toprule
Method & Bounds & Objective & BND & BYLD & Bond total & Largest ETF & HHI & Gross & Short \\
\midrule
Historical & LO & MC95 & 8.57 & 3.33 & 11.90 & 8.57 & 0.0388 & 100.00 & 0.00 \\
Historical & LS & MC95 & 8.65 & 3.33 & 11.98 & 8.65 & 0.0379 & 101.82 & 0.91 \\
Dynamic & LO & MC95 & 8.37 & 3.38 & 11.76 & 8.37 & 0.0376 & 100.00 & 0.00 \\
Dynamic & LS & MC95 & 8.37 & 3.38 & 11.76 & 8.37 & 0.0376 & 100.00 & 0.00 \\
\bottomrule\end{tabular}}
\end{table}
The MC95 strategies had similar ETF-level concentration and direct holdings in BND and BYLD (Table~\ref{tab:allocation_reconciled_mc95_2026}). The greater short exposure of the historical long--short portfolio contrasts with the conditional portfolio's effectively long-only positions. The comparison is consistent with the view that realized exposures depended on both the estimated inputs and the opportunity to change weights under the restrictive adjustment rule. It does not describe look-through fixed-income exposure inside other ETFs.

\FloatBarrier

\subsection{Sensitivity analyses}\label{sec:sensitivity2026}
We examined how portfolio outcomes varied with the permitted change in target weights, the simulated scenario count and random seed, and the length of the moving estimation window. Each comparison retained the ETF universe and portfolio objectives and used common evaluation dates within that experiment. Table~\ref{tab:sensitivity_summary2026} summarizes the observed patterns; Appendix~\ref{app:sensitivity2026} gives the strategy-level results. These descriptive exercises were not independent tests of statistical superiority.

\begin{table}[htbp]\centering\small
\caption{Overview of sensitivity experiments. Changes are descriptive and were not subject to a separate multiplicity-adjusted test.}\label{tab:sensitivity_summary2026}
\begin{tabularx}{\textwidth}{@{}p{0.23\textwidth}p{0.23\textwidth}X@{}}\toprule
Experiment & Alternatives & Main observation \\ \midrule
Daily target-turnover cap & $1/(30\times252)$ versus $0.01$ & Under the 1\% cap, concentration and implementation turnover rise across the examined portfolios; minimum-risk realized return and risk can both decline, whereas selected tangent portfolios have higher gross returns. \\
Dynamic scenario count & 5,000, 10,000 and 20,000 & Annualized geometric-return changes from the 10,000-scenario baseline are small under the restrictive baseline turnover cap. \\
Random seed & 20261015 versus 12345 & Historical portfolio returns are unchanged; dynamic annualized-return differences are small in the two-seed comparison. \\
Moving estimation window & 756, 1008 and 1260 & The shorter window generally produces lower matched annualized returns; longer-window effects are mixed across strategies. \\
\bottomrule\end{tabularx}\end{table}

\begin{figure}[htbp]\centering
\includegraphics[width=\textwidth]{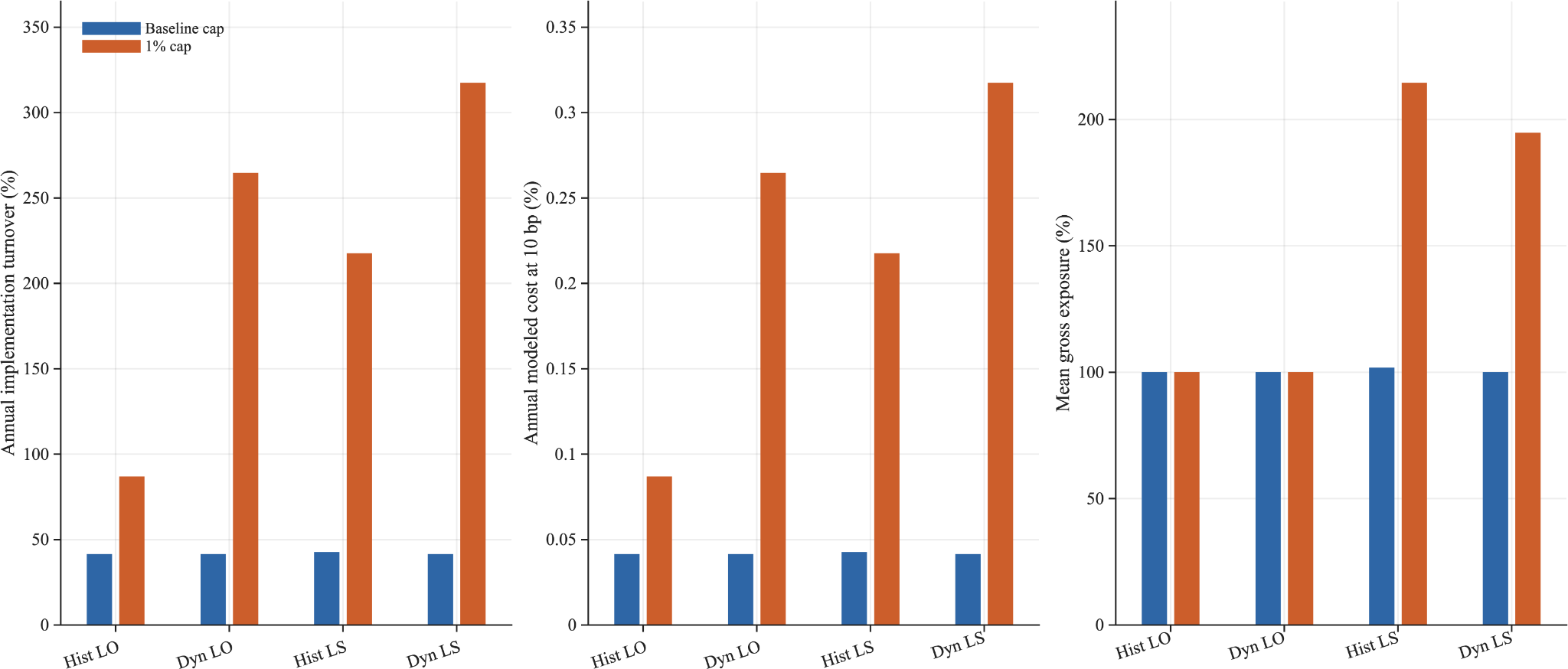}
\caption{MC95 turnover-cap sensitivity for historical and dynamic LO/LS portfolios. Bars compare annualized implementation turnover (left), annual modeled costs under a 10-basis-point linear cost rate (middle), and average gross exposure (right) for the baseline and 1\% daily target-adjustment caps. All plotted quantities are percentages; a larger turnover allowance does not guarantee a higher net return.}\label{fig:cap_sensitivity_mc95_2026}
\end{figure}

Allowing larger daily changes in target weights had a more visible effect on the portfolios than changing the simulation settings. With greater freedom to move toward their selected allocations, many strategies traded more and held more concentrated exposures. Several minimum-risk strategies experienced lower realized returns alongside lower measured risk, whereas some tangent strategies earned higher gross returns. The effect was not uniformly favorable to either return-input method. Figure~\ref{fig:cap_sensitivity_mc95_2026} shows the accompanying changes in implementation turnover, modeled costs, and gross exposure for MC95 portfolios; Appendix Table~\ref{tab:sensitivity_turnover_all} reports the portfolio-level gross and net returns. The cost estimates retain the same linear trading assumption and exclude initial formation, market impact, and short-position financing.

In contrast, varying the conditional scenario count around its baseline setting led to relatively small changes in realized returns (Appendix Table~\ref{tab:sensitivity_scenarios_all}). Changing the random seed produced a similarly limited response in the conditional portfolios, while historical optimization was unaffected (Appendix Table~\ref{tab:sensitivity_seed_all}). With a restrictive adjustment rule, changes in estimated optimal weights could affect actual holdings only gradually; this offers one explanation for the modest sensitivity to simulation settings. These results do not establish convergence of simulated tail losses or of the unconstrained optimal portfolio.

Changing the moving estimation window gave a less uniform pattern (Appendix Table~\ref{tab:sensitivity_window_all}). The shorter window generally had lower matched realized returns, while the longer-window results varied by portfolio objective and investment bounds. A shorter history emphasizes more recent observations but offers less data for estimating volatility and dependence; a longer history changes that balance. The experiment does not identify a uniformly preferable window length. Nor do the one-at-a-time tests cover every combination of trading costs, rebalancing schedules, marginal models, or dependence specifications.
\FloatBarrier

\section{Discussion}\label{sec:discussion}

The matched comparison shows why differences in estimated return distributions need not produce similarly large differences in realized portfolio outcomes. The historical and conditional frontiers displayed different estimated return--risk opportunities, yet the restrictive adjustment rule often kept their subsequent holdings close. This finding does not establish that the underlying forecasts were equally accurate: it shows how a common trading restriction shaped the portfolios that could be constructed from them.

Historical inputs generally had higher observed reward--risk ratios, especially in several tangent strategies (Table~\ref{tab:rr2026}). Differences in downside risk were less consistent. A minimum-risk objective can favor holdings with lower modeled loss exposure but also less participation in positive market returns, while a tangent objective seeks a higher expected excess return relative to a risk measure. The outcomes in Tables~\ref{tab:riskmed2026} and \ref{tab:grossnet2026} are consistent with these portfolio-objective tradeoffs. They do not imply that historical returns provided better forecasts in general.

The paired bootstrap results also limit the inference. Some unadjusted intervals excluded zero, but none of the comparisons remained significant after adjustment for multiple testing. The absence of adjusted rejections is not evidence that the methods are equivalent. Moreover, the bootstrap used the realized return series from already-optimized portfolios rather than repeating the full estimation and portfolio-selection exercise within each resample. Its uncertainty statements therefore apply to the observed strategy histories under the stated resampling design.

Implementation adds a separate consideration. The baseline rule limited changes between successive target weights, but market movements continually altered the proportions actually held. Maintaining the targets could therefore require appreciable rebalancing even while target adjustments were small (Table~\ref{tab:turn2026} and Figure~\ref{fig:turnover_paths_mc95_2026}). When the adjustment allowance was relaxed, portfolio concentration, short exposure, turnover, and performance responded differently across objectives. The return-input comparison consequently depended on the investment rule used to translate estimated preferences into holdings.

The sensitivity results support this interpretation but do not isolate a single causal mechanism. Changing scenario counts or seeds had relatively little effect under the baseline cap, whereas the effect of a looser cap was more substantial for several strategies; changing the estimation window produced mixed outcomes. These comparisons concern the tested settings and sample, and should not be extended to other trading restrictions or models without further evaluation.

Several limitations matter for investment use. The Yahoo Finance price adjustments were not independently reconstructed. Reported returns used next-session, close-to-close attribution rather than observed executable trades, and transaction costs were modeled without market impact, variable spreads, or borrowing and financing costs. The conditional copula permitted joint-tail dependence but was not tested for predictive loss calibration. Numerical checks verified feasibility and stability at representative decisions, not global optimality across the full path. The single ETF universe and evaluation period further limit generalization.

\section{Conclusions}\label{sec:conclusions}

This paper compared historical moving-window returns with conditional scenario inputs for multi-asset ETF optimization under matched portfolio objectives and investment rules. Historical portfolios generally had higher realized reward--risk ratios over the evaluation period, but none of the paired differences remained significant after adjustment for multiple testing. Conditional scenarios did not consistently improve realized downside-risk measures. These findings describe outcomes under the specified data and trading conventions, rather than a general ranking of forecasting methods.

The portfolio implementation results offer the main practical interpretation. A tight constraint on changes between target allocations limited the effects of different return inputs on realized holdings, while maintaining those allocations still required trades as market prices changed. Relaxing that constraint altered trading activity, exposures, and some realized returns. For evaluating portfolio risk management, the choice of return inputs should therefore be considered together with the limits and costs of implementing the resulting allocations.

\authorcontributions{Conceptualization, A.A.M., W.B.L., S.T.R. and F.J.F.; methodology, A.A.M., A.J., W.B.L. and S.T.R.; software, A.A.M. and N.A.; validation, A.A.M., A.J. and N.A.; formal analysis, A.A.M., A.J. and N.A.; investigation, A.A.M., A.J. and N.A.; resources, S.T.R. and F.J.F.; data curation, A.A.M.; writing---original draft preparation, A.A.M. and N.A.; writing---review and editing, A.J., N.A., W.B.L., S.T.R. and F.J.F.; visualization, A.A.M.; supervision, W.B.L., S.T.R. and F.J.F.; project administration, S.T.R. and F.J.F.}
\funding{This research received no external funding.}
\institutionalreview{Not applicable.}
\informedconsent{Not applicable.}
\dataavailability{The ETF adjusted closing-price series are obtained from Yahoo Finance. The source series are subject to the data provider's terms. Derived portfolio statistics, supplementary results and analysis code may be requested from the corresponding author, subject to applicable sharing permissions.}
\conflictsofinterest{The authors declare no conflicts of interest.}

\appendixstart
\appendix
\section{Data and Supplementary Materials}
\label{app:data}
Table~\ref{tab:ETFList} records the ETF universe used in the analysis. ETF categories are descriptive classifications and not look-through estimates of fund asset allocations.
\begin{table}[htbp]\centering\small
\caption{The 30 U.S.-listed ETFs used in the analysis.}\label{tab:ETFList}
\begin{tabular}{lll}\toprule Ticker & ETF & Broad category \\
\midrule
VT & Vanguard Total World Stock & Global equity \\
VTI & Vanguard Total Stock Market & U.S. equity \\
ACWI & iShares MSCI ACWI & Global equity \\
ACWX & iShares MSCI ACWI ex U.S. & International equity \\
SUSA & iShares ESG Optimized MSCI USA & U.S. equity \\
ROUS & Hartford Multifactor U.S. Equity & U.S. equity factor \\
IYG & iShares U.S. Financials & Sector equity \\
BND & Vanguard Total Bond Market & Fixed income \\
FPE & First Trust Preferred Securities and Income & Preferred income \\
BYLD & iShares Yield Optimized Bond & Fixed income \\
NTSX & WisdomTree U.S. Efficient Core & Multi-asset equity/bond \\
GAL & SPDR Global Allocation & Multi-asset allocation \\
MDIV & First Trust Multi-Asset Diversified Income & Multi-asset income \\
AOM & iShares Core Moderate Allocation & Multi-asset allocation \\
AOR & iShares Core Growth Allocation & Multi-asset allocation \\
INKM & SPDR Income Allocation & Multi-asset income \\
HNDL & Strategy Shares Nasdaq 7HANDL Index & Multi-asset income \\
ALTY & Global X SuperDividend Alternatives & Alternatives \\
IYLD & iShares Morningstar Multi-Asset Income & Multi-asset income \\
CVY & Invesco Zacks Multi-Asset Income & Multi-asset income \\
PDBC & Invesco Optimum Yield Diversified Commodity & Commodities \\
QAI & IQ Hedge Multi-Strategy & Alternatives \\
HDG & ProShares Hedge Replication & Alternatives \\
MNA & IQ Merger Arbitrage & Alternatives \\
RAAX & VanEck Real Asset Allocation & Real assets \\
RLY & SPDR Multi-Asset Real Return & Real assets \\
SRET & Global X SuperDividend REIT & Real estate \\
YYY & Amplify High Income & Income \\
GAA & Cambria Global Asset Allocation & Multi-asset allocation \\
GMOM & Cambria Global Momentum & Tactical allocation \\
\bottomrule\end{tabular}
\end{table}

\section{ETF Return Correlations}\label{app:correlations2026}
\begin{figure}[htbp]
\centering
\includegraphics[width=0.80\textwidth]{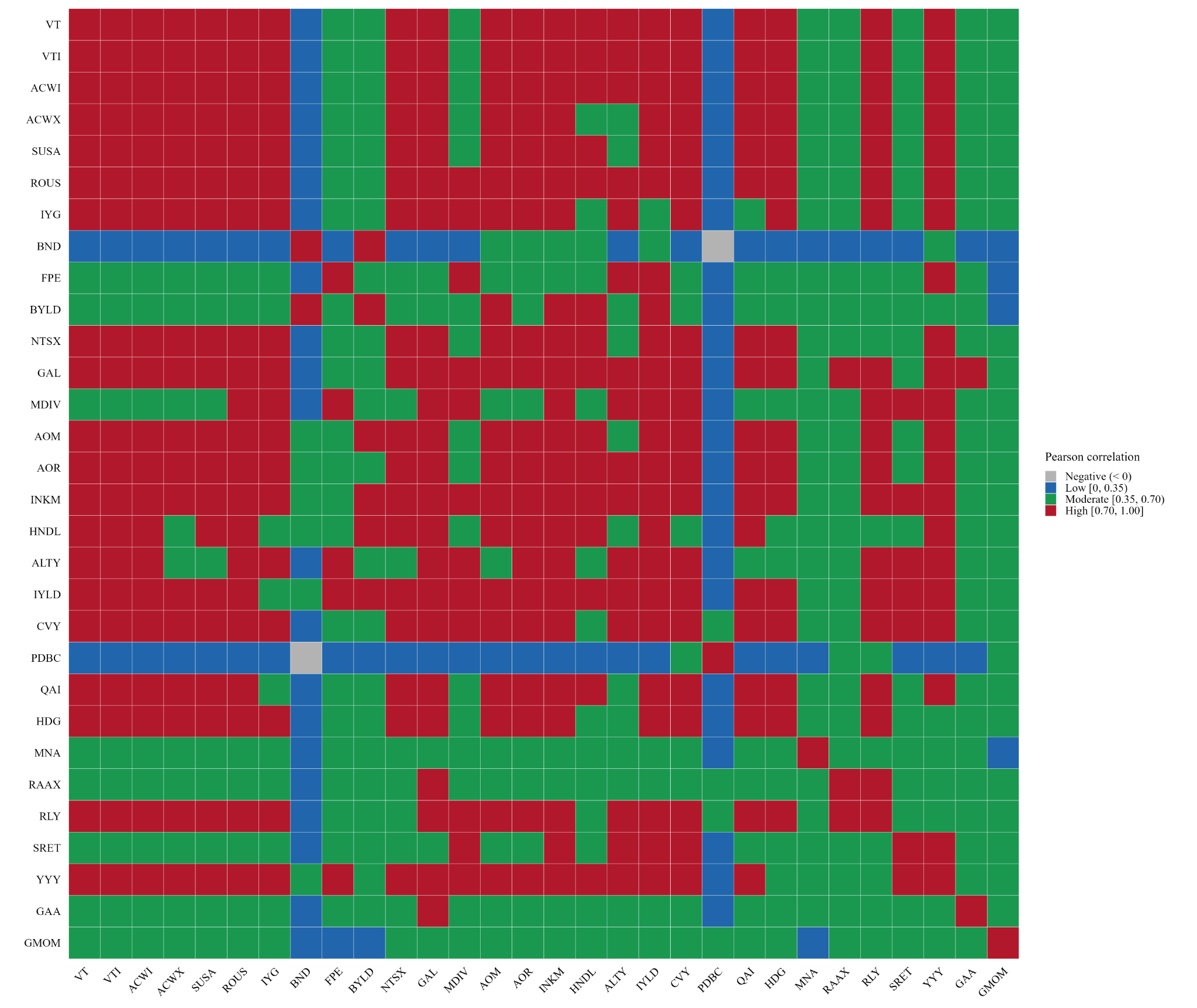}
\caption{Pairwise Pearson correlations of daily arithmetic returns for the 30 U.S.-listed ETFs, 20 December 2018--30 September 2026 (1,954 common return observations). Color categories denote negative ($r<0$), low ($0\leq r<0.35$), moderate ($0.35\leq r<0.70$), and high ($0.70\leq r\leq1$) sample correlations. These bands are descriptive, not statistical significance tests.}
\label{fig:etf_correlation_2026}
\end{figure}

Appendix Figure~\ref{fig:etf_correlation_2026} presents the return-correlation matrix.
\FloatBarrier

\section{Indexed Adjusted-Price Paths}\label{app:pricepaths2026}
\begin{figure}[htbp]
\centering
\includegraphics[width=0.85\textwidth,height=0.76\textheight,keepaspectratio]{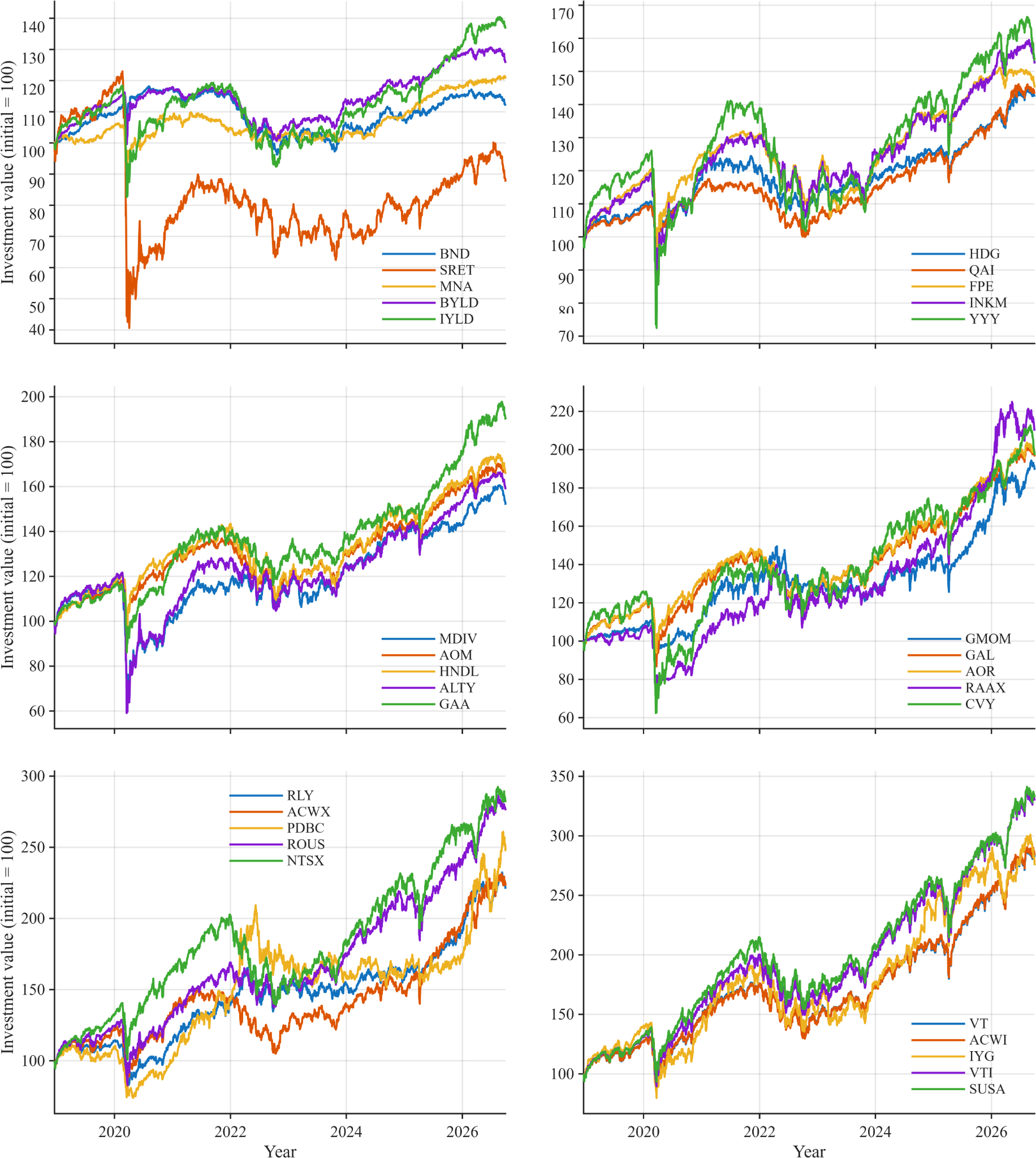}
\caption{Indexed adjusted closing-price paths for the 30 ETFs, 19 December 2018--30 September 2026. Each ETF series is normalized to 100 at its initial sample price. The ETFs are sorted by their full-sample mean daily arithmetic return and displayed in six successive groups of five. Panel scales differ to make within-group paths visible. These are descriptive normalized adjusted-price paths, not the gross or net out-of-sample optimized portfolio wealth paths.}
\label{fig:etf_indexed_prices_2026}
\end{figure}

Appendix Figure~\ref{fig:etf_indexed_prices_2026} displays the adjusted-price paths.
\FloatBarrier

\section{Numerical Validation and Simulation Diagnostics}\label{app:numerical_checks}
The baseline audit reconstructed the stored portfolio weights and realized gross returns for all 22,704 daily strategy decisions: 11,352 historical and 11,352 conditional. The greatest budget residual was $2.72\times10^{-10}$; no position bounds were violated, and the largest positive excess above the target-adjustment cap was $6.75\times10^{-18}$. The maximum absolute difference between a stored realized gross return and its value recomputed from the weights and observed ETF returns was $1.39\times10^{-17}$. All decisions passed the production acceptance checks. These feasibility diagnostics are not certificates of numerical optimality.

For numerical cap overshoots of at most $10^{-6}$, candidates were moved along the line joining the candidate and the previous feasible target until the cap was met. The optimizer was not given a larger turnover allowance. Of 3,983 historical and 3,774 conditional strategy decisions flagged for such a correction, 7,659 of the combined 7,757 flags (98.74\%) had a recorded maximum candidate overshoot at most $10^{-15}$; 66 exceeded $10^{-8}$, and the maximum was $9.92\times10^{-8}$. In tangent-CVaR optimization, each decision-level flag indicated that at least one frontier candidate was corrected, not necessarily the ultimately selected allocation. Corrected candidate returns and CVaR values were recalculated before the reward--risk ratio was compared. The original frontier target-return levels were not preserved in the production logs; post-correction satisfaction of those original auxiliary target-return constraints could not be verified independently.

The tangent-CVaR grid check used 20, 50, and 100 requested frontier candidates under identical historical inputs or identical reconstructed 10,000-scenario conditional inputs and the same preceding target allocation. The first evaluated decision, 21~December~2022, covered TC95 and TC99 under LO and LS bounds for both input methods. The 20-candidate replays agreed with the archived baseline portfolio weights, with maximum one-way distance $1.56\times10^{-15}$. The largest relative change in the historical excess-return-to-CVaR ratio between the 20-point and finer grids was 0.000231\%, and the maximum one-way allocation difference was $1.30\times10^{-5}$. Conditional selected portfolios did not change to recorded precision. Tightening the CVaR solver's absolute and relative gap settings from $10^{-8}$ to $10^{-10}$ did not change the first-date selected allocations or ratios at recorded precision.

A further test of eight historical flagged tangent-CVaR decisions, including four with candidate overshoots exceeding $10^{-8}$, produced a largest absolute ratio change of $1.74\times10^{-7}$ (0.00136\% relative) and a maximum one-way distance of $2.95\times10^{-5}$ from the replayed 20-point allocation. All checked final weights satisfied the budget, bounds, and target-adjustment acceptance tests. These results describe stability at the tested dates only; the complete 946-date optimization paths were not rerun at alternative grid resolutions.

For scenario reproducibility, ARMA(1,1)--GARCH(1,1) models with Student-$t$ estimation likelihoods were fitted by MATLAB's \texttt{estimate} subject to the toolbox's admissibility restrictions; no bespoke coefficient bounds were imposed. A failed marginal fit after the initial window would have led to reuse of that ETF's preceding fitted model, but no such reuse appeared in the archived diagnostics. The standardized residual distribution was estimated with \texttt{ksdensity} using its default kernel and automatic bandwidth; the saved bandwidths ranged from 0.179447 to 0.269415. A Student-$t$ copula was fitted with \texttt{copulafit} using approximate maximum likelihood, followed by \texttt{copularnd} and kernel-inverse simulation. The Mersenne Twister was seeded once at 20261015, and simulated innovations were neither recentered nor rescaled. Across saved ETF--date diagnostics, their sample means ranged from $-0.165013$ to $0.034645$ and their standard deviations from $0.828877$ to $1.289241$. These records document implementation choices and numerical behavior; they do not validate predictive tail calibration.

\section{Individual Bootstrap $p$-Values}\label{app:bootstrap_pvalues2026}
For each matched portfolio and ratio, let $\widehat\Delta$ denote the observed dynamic-minus-historical difference and let $\widehat\Delta^{*(b)}$, $b=1,\ldots,B$, denote its paired stationary-bootstrap replicate. The reported centered two-sided Monte Carlo $p$-value is
\begin{equation}\label{eq:bootstrap_centered_p2026}
\widehat p=\frac{1+\sum_{b=1}^{B}\mathbf{1}\!\left\{\left|\widehat\Delta^{*(b)}-\widehat\Delta\right|\geq\left|\widehat\Delta\right|\right\}}{B+1},\qquad B=5{,}000.
\end{equation}
The Holm and Benjamini--Hochberg procedures \citep{Holm1979,BenjaminiHochberg1995} are applied separately to each family of 48 raw $p$-values (one family per expected block length). Tables~\ref{tab:bootstrap_pvals_LO_2026} and \ref{tab:bootstrap_pvals_LS_2026} contain the complete reference-length ($L=20$) results. Tables~\ref{tab:bootstrap_alt_10_LO_2026}--\ref{tab:bootstrap_alt_40_LS_2026} report the remaining individual comparisons and intervals at $L=10$ and $L=40$. The individual 95\% percentile confidence intervals in Table~\ref{tab:paired_bootstrap_detailed_2026} and the centered-bootstrap $p$-values are different inferential constructions and need not yield identical rejection decisions.

\begin{table}[p]
\centering\scriptsize
\caption{Individual two-sided centered-bootstrap $p$-values for long--only portfolios, expected block length $L=20$ and $B=5{,}000$ paired resamples. The Holm and BH columns are multiplicity-adjusted across all 48 comparisons, including both bounds.}
\label{tab:bootstrap_pvals_LO_2026}
\setlength{\tabcolsep}{5pt}
\begin{tabular}{llrrr}\toprule
Portfolio & Ratio & Raw $p$ & Holm $p$ & BH $p$ \\
\midrule
MV & Rachev 95/95 & 0.4507 & 1.0000 & 0.6181 \\
MV & SS$_2$ & 0.1160 & 1.0000 & 0.3767 \\
MV & STAR 95 & 0.1554 & 1.0000 & 0.4143 \\
MV & Sharpe & 0.2250 & 1.0000 & 0.4695 \\
TV & Rachev 95/95 & 0.3333 & 1.0000 & 0.5161 \\
TV & SS$_2$ & 0.8906 & 1.0000 & 0.9096 \\
TV & STAR 95 & 0.5423 & 1.0000 & 0.7231 \\
TV & Sharpe & 0.4495 & 1.0000 & 0.6181 \\
MC95 & Rachev 95/95 & 0.2753 & 1.0000 & 0.4720 \\
MC95 & SS$_2$ & 0.0390 & 1.0000 & 0.3407 \\
MC95 & STAR 95 & 0.1214 & 1.0000 & 0.3767 \\
MC95 & Sharpe & 0.1420 & 1.0000 & 0.4009 \\
TC95 & Rachev 95/95 & 0.5819 & 1.0000 & 0.7456 \\
TC95 & SS$_2$ & 0.8742 & 1.0000 & 0.9096 \\
TC95 & STAR 95 & 0.2679 & 1.0000 & 0.4720 \\
TC95 & Sharpe & 0.2150 & 1.0000 & 0.4695 \\
MC99 & Rachev 95/95 & 0.7103 & 1.0000 & 0.8243 \\
MC99 & SS$_2$ & 0.0484 & 1.0000 & 0.3407 \\
MC99 & STAR 95 & 0.0546 & 1.0000 & 0.3407 \\
MC99 & Sharpe & 0.0550 & 1.0000 & 0.3407 \\
TC99 & Rachev 95/95 & 0.6649 & 1.0000 & 0.8183 \\
TC99 & SS$_2$ & 0.8818 & 1.0000 & 0.9096 \\
TC99 & STAR 95 & 0.2679 & 1.0000 & 0.4720 \\
TC99 & Sharpe & 0.2188 & 1.0000 & 0.4695 \\
\bottomrule\end{tabular}
\end{table}
\FloatBarrier
\begin{table}[p]
\centering\scriptsize
\caption{Individual two-sided centered-bootstrap $p$-values for limited long--short portfolios, expected block length $L=20$ and $B=5{,}000$ paired resamples. The Holm and BH columns are multiplicity-adjusted across all 48 comparisons, including both bounds.}
\label{tab:bootstrap_pvals_LS_2026}
\setlength{\tabcolsep}{5pt}
\begin{tabular}{llrrr}\toprule
Portfolio & Ratio & Raw $p$ & Holm $p$ & BH $p$ \\
\midrule
MV & Rachev 95/95 & 0.1256 & 1.0000 & 0.3767 \\
MV & SS$_2$ & 0.0138 & 0.6623 & 0.3407 \\
MV & STAR 95 & 0.0232 & 1.0000 & 0.3407 \\
MV & Sharpe & 0.0250 & 1.0000 & 0.3407 \\
TV & Rachev 95/95 & 0.5903 & 1.0000 & 0.7456 \\
TV & SS$_2$ & 0.7740 & 1.0000 & 0.8641 \\
TV & STAR 95 & 0.3533 & 1.0000 & 0.5300 \\
TV & Sharpe & 0.3309 & 1.0000 & 0.5161 \\
MC95 & Rachev 95/95 & 0.2937 & 1.0000 & 0.4862 \\
MC95 & SS$_2$ & 0.0568 & 1.0000 & 0.3407 \\
MC95 & STAR 95 & 0.0876 & 1.0000 & 0.3767 \\
MC95 & Sharpe & 0.1002 & 1.0000 & 0.3767 \\
TC95 & Rachev 95/95 & 0.9226 & 1.0000 & 0.9226 \\
TC95 & SS$_2$ & 0.7213 & 1.0000 & 0.8243 \\
TC95 & STAR 95 & 0.2653 & 1.0000 & 0.4720 \\
TC95 & Sharpe & 0.2198 & 1.0000 & 0.4695 \\
MC99 & Rachev 95/95 & 0.4467 & 1.0000 & 0.6181 \\
MC99 & SS$_2$ & 0.1142 & 1.0000 & 0.3767 \\
MC99 & STAR 95 & 0.0908 & 1.0000 & 0.3767 \\
MC99 & Sharpe & 0.1004 & 1.0000 & 0.3767 \\
TC99 & Rachev 95/95 & 0.8298 & 1.0000 & 0.9053 \\
TC99 & SS$_2$ & 0.7091 & 1.0000 & 0.8243 \\
TC99 & STAR 95 & 0.2631 & 1.0000 & 0.4720 \\
TC99 & Sharpe & 0.2234 & 1.0000 & 0.4695 \\
\bottomrule\end{tabular}
\end{table}
\FloatBarrier

\FloatBarrier

\section{Bootstrap Block-Length Robustness}\label{app:bootstrap2026}
Table~\ref{tab:bootstrap2026} reports sensitivity of the paired bootstrap results to block length.
\begin{table}[htbp]\centering
\scriptsize
\caption{Paired stationary-bootstrap robustness to expected block length. Each specification comprises 48 paired portfolio--metric comparisons.}\label{tab:bootstrap2026}
\setlength{\tabcolsep}{3pt}
\begin{tabular}{rrrrr}\toprule

Expected block length & Resamples & 95\% CIs excluding zero & Holm (5\%) & BH (5\%) \\
\midrule

10 & 5000 & 5 & 0 & 0 \\

20 & 5000 & 6 & 0 & 0 \\

40 & 5000 & 8 & 0 & 0 \\

\bottomrule\end{tabular}\end{table}
\FloatBarrier

\section{Individual Bootstrap Results at Alternative Block Lengths}\label{app:bootstrap_alternative_blocks}
The following tables report the individual comparisons at expected block lengths of 10 and 40 trading sessions. The corresponding block-length-20 results appear in Tables~\ref{tab:bootstrap_pvals_LO_2026} and \ref{tab:bootstrap_pvals_LS_2026} and in the confidence-interval comparisons in Table~\ref{tab:paired_bootstrap_detailed_2026}. Each interval is an unadjusted percentile interval; Holm and Benjamini--Hochberg corrections were applied to the 48 raw $p$-values within the stated block length.
\begin{table}[p]
\centering\scriptsize
\caption{Individual paired stationary-bootstrap differences, unadjusted 95\% percentile confidence intervals, and centered-bootstrap $p$-values for long--only portfolios, expected block length $L=10$ ($B=5{,}000$). Adjustments were calculated within the complete family of 48 comparisons.}
\label{tab:bootstrap_alt_10_LO_2026}
\setlength{\tabcolsep}{2.7pt}
\resizebox{\textwidth}{!}{%
\begin{tabular}{llrrrrrr}\toprule
Portfolio & Ratio & Difference & CI lower & CI upper & Raw $p$ & Holm $p$ & BH $p$ \\
\midrule
MV & Rachev 95/95 & -0.001131 & -0.004033 & 0.002152 & 0.4513 & 1.0000 & 0.6299 \\
MV & SS$_2$ & -0.000824 & -0.001849 & 0.000298 & 0.1292 & 1.0000 & 0.3817 \\
MV & STAR 95 & -0.000209 & -0.000515 & 0.000096 & 0.1824 & 1.0000 & 0.4823 \\
MV & Sharpe & -0.000373 & -0.001032 & 0.000254 & 0.2599 & 1.0000 & 0.4823 \\
TV & Rachev 95/95 & 0.006675 & -0.006956 & 0.020389 & 0.3377 & 1.0000 & 0.5229 \\
TV & SS$_2$ & -0.000412 & -0.007077 & 0.005029 & 0.8836 & 1.0000 & 0.9024 \\
TV & STAR 95 & -0.000679 & -0.003030 & 0.001423 & 0.5407 & 1.0000 & 0.7209 \\
TV & Sharpe & -0.001887 & -0.007172 & 0.002737 & 0.4487 & 1.0000 & 0.6299 \\
MC95 & Rachev 95/95 & -0.003544 & -0.010229 & 0.002909 & 0.2813 & 1.0000 & 0.4823 \\
MC95 & SS$_2$ & -0.002394 & -0.004891 & -0.000282 & 0.0420 & 1.0000 & 0.3719 \\
MC95 & STAR 95 & -0.000627 & -0.001449 & 0.000128 & 0.1116 & 1.0000 & 0.3817 \\
MC95 & Sharpe & -0.001272 & -0.003005 & 0.000395 & 0.1352 & 1.0000 & 0.3817 \\
TC95 & Rachev 95/95 & 0.004422 & -0.012248 & 0.019753 & 0.5787 & 1.0000 & 0.7507 \\
TC95 & SS$_2$ & -0.000605 & -0.009543 & 0.005395 & 0.8732 & 1.0000 & 0.9024 \\
TC95 & STAR 95 & -0.001203 & -0.003566 & 0.000774 & 0.2721 & 1.0000 & 0.4823 \\
TC95 & Sharpe & -0.002908 & -0.007849 & 0.001233 & 0.2086 & 1.0000 & 0.4823 \\
MC99 & Rachev 95/95 & -0.001529 & -0.009484 & 0.007900 & 0.7265 & 1.0000 & 0.8302 \\
MC99 & SS$_2$ & -0.002664 & -0.005429 & 0.000067 & 0.0572 & 1.0000 & 0.3719 \\
MC99 & STAR 95 & -0.000843 & -0.001738 & 0.000040 & 0.0594 & 1.0000 & 0.3719 \\
MC99 & Sharpe & -0.001778 & -0.003663 & 0.000025 & 0.0596 & 1.0000 & 0.3719 \\
TC99 & Rachev 95/95 & 0.003528 & -0.013053 & 0.018949 & 0.6573 & 1.0000 & 0.8089 \\
TC99 & SS$_2$ & -0.000575 & -0.009464 & 0.005366 & 0.8810 & 1.0000 & 0.9024 \\
TC99 & STAR 95 & -0.001188 & -0.003505 & 0.000756 & 0.2751 & 1.0000 & 0.4823 \\
TC99 & Sharpe & -0.002823 & -0.007703 & 0.001251 & 0.2132 & 1.0000 & 0.4823 \\
\bottomrule\end{tabular}}
\end{table}
\FloatBarrier

\begin{table}[p]
\centering\scriptsize
\caption{Individual paired stationary-bootstrap differences, unadjusted 95\% percentile confidence intervals, and centered-bootstrap $p$-values for limited long--short portfolios, expected block length $L=10$ ($B=5{,}000$). Adjustments were calculated within the complete family of 48 comparisons.}
\label{tab:bootstrap_alt_10_LS_2026}
\setlength{\tabcolsep}{2.7pt}
\resizebox{\textwidth}{!}{%
\begin{tabular}{llrrrrrr}\toprule
Portfolio & Ratio & Difference & CI lower & CI upper & Raw $p$ & Holm $p$ & BH $p$ \\
\midrule
MV & Rachev 95/95 & -0.000298 & -0.000619 & 0.000090 & 0.1112 & 1.0000 & 0.3817 \\
MV & SS$_2$ & -0.000164 & -0.000269 & -0.000031 & 0.0102 & 0.4895 & 0.3423 \\
MV & STAR 95 & -0.000038 & -0.000069 & -0.000006 & 0.0204 & 0.9586 & 0.3423 \\
MV & Sharpe & -0.000076 & -0.000140 & -0.000011 & 0.0214 & 0.9842 & 0.3423 \\
TV & Rachev 95/95 & -0.003698 & -0.017695 & 0.010732 & 0.6017 & 1.0000 & 0.7600 \\
TV & SS$_2$ & -0.000990 & -0.008959 & 0.005126 & 0.7624 & 1.0000 & 0.8511 \\
TV & STAR 95 & -0.001120 & -0.003588 & 0.001149 & 0.3509 & 1.0000 & 0.5264 \\
TV & Sharpe & -0.002570 & -0.008086 & 0.002266 & 0.3229 & 1.0000 & 0.5167 \\
MC95 & Rachev 95/95 & -0.004211 & -0.012354 & 0.004559 & 0.3053 & 1.0000 & 0.5054 \\
MC95 & SS$_2$ & -0.002847 & -0.006034 & -0.000039 & 0.0620 & 1.0000 & 0.3719 \\
MC95 & STAR 95 & -0.000903 & -0.001980 & 0.000075 & 0.0826 & 1.0000 & 0.3817 \\
MC95 & Sharpe & -0.001852 & -0.004135 & 0.000253 & 0.0974 & 1.0000 & 0.3817 \\
TC95 & Rachev 95/95 & -0.000963 & -0.020063 & 0.016337 & 0.9092 & 1.0000 & 0.9092 \\
TC95 & SS$_2$ & -0.001464 & -0.011176 & 0.005349 & 0.7205 & 1.0000 & 0.8302 \\
TC95 & STAR 95 & -0.001385 & -0.004022 & 0.000875 & 0.2679 & 1.0000 & 0.4823 \\
TC95 & Sharpe & -0.003228 & -0.008860 & 0.001610 & 0.2184 & 1.0000 & 0.4823 \\
MC99 & Rachev 95/95 & -0.003535 & -0.013011 & 0.006603 & 0.4593 & 1.0000 & 0.6299 \\
MC99 & SS$_2$ & -0.002463 & -0.005981 & 0.000511 & 0.1316 & 1.0000 & 0.3817 \\
MC99 & STAR 95 & -0.000874 & -0.001944 & 0.000140 & 0.0980 & 1.0000 & 0.3817 \\
MC99 & Sharpe & -0.001795 & -0.004059 & 0.000345 & 0.1034 & 1.0000 & 0.3817 \\
TC99 & Rachev 95/95 & -0.002046 & -0.021137 & 0.015261 & 0.8248 & 1.0000 & 0.8998 \\
TC99 & SS$_2$ & -0.001514 & -0.011011 & 0.005153 & 0.7095 & 1.0000 & 0.8302 \\
TC99 & STAR 95 & -0.001359 & -0.003980 & 0.000878 & 0.2675 & 1.0000 & 0.4823 \\
TC99 & Sharpe & -0.003114 & -0.008604 & 0.001637 & 0.2232 & 1.0000 & 0.4823 \\
\bottomrule\end{tabular}}
\end{table}
\FloatBarrier

\begin{table}[p]
\centering\scriptsize
\caption{Individual paired stationary-bootstrap differences, unadjusted 95\% percentile confidence intervals, and centered-bootstrap $p$-values for long--only portfolios, expected block length $L=40$ ($B=5{,}000$). Adjustments were calculated within the complete family of 48 comparisons.}
\label{tab:bootstrap_alt_40_LO_2026}
\setlength{\tabcolsep}{2.7pt}
\resizebox{\textwidth}{!}{%
\begin{tabular}{llrrrrrr}\toprule
Portfolio & Ratio & Difference & CI lower & CI upper & Raw $p$ & Holm $p$ & BH $p$ \\
\midrule
MV & Rachev 95/95 & -0.001131 & -0.003731 & 0.001957 & 0.4073 & 1.0000 & 0.5925 \\
MV & SS$_2$ & -0.000824 & -0.001834 & 0.000170 & 0.1060 & 1.0000 & 0.3634 \\
MV & STAR 95 & -0.000209 & -0.000485 & 0.000071 & 0.1350 & 1.0000 & 0.3811 \\
MV & Sharpe & -0.000373 & -0.000970 & 0.000182 & 0.1974 & 1.0000 & 0.4401 \\
TV & Rachev 95/95 & 0.006675 & -0.006645 & 0.020316 & 0.3257 & 1.0000 & 0.5044 \\
TV & SS$_2$ & -0.000412 & -0.007882 & 0.005061 & 0.8950 & 1.0000 & 0.9034 \\
TV & STAR 95 & -0.000679 & -0.003130 & 0.001218 & 0.5277 & 1.0000 & 0.6878 \\
TV & Sharpe & -0.001887 & -0.007602 & 0.002371 & 0.4441 & 1.0000 & 0.6091 \\
MC95 & Rachev 95/95 & -0.003544 & -0.009794 & 0.002779 & 0.2561 & 1.0000 & 0.4401 \\
MC95 & SS$_2$ & -0.002394 & -0.004834 & -0.000391 & 0.0352 & 1.0000 & 0.2711 \\
MC95 & STAR 95 & -0.000627 & -0.001502 & 0.000164 & 0.1318 & 1.0000 & 0.3811 \\
MC95 & Sharpe & -0.001272 & -0.003083 & 0.000451 & 0.1548 & 1.0000 & 0.4127 \\
TC95 & Rachev 95/95 & 0.004422 & -0.009716 & 0.019341 & 0.5429 & 1.0000 & 0.6878 \\
TC95 & SS$_2$ & -0.000605 & -0.009869 & 0.005144 & 0.8778 & 1.0000 & 0.9034 \\
TC95 & STAR 95 & -0.001203 & -0.003560 & 0.000601 & 0.2535 & 1.0000 & 0.4401 \\
TC95 & Sharpe & -0.002908 & -0.008113 & 0.001056 & 0.2126 & 1.0000 & 0.4401 \\
MC99 & Rachev 95/95 & -0.001529 & -0.008977 & 0.007803 & 0.7065 & 1.0000 & 0.8346 \\
MC99 & SS$_2$ & -0.002664 & -0.005142 & -0.000365 & 0.0338 & 1.0000 & 0.2711 \\
MC99 & STAR 95 & -0.000843 & -0.001652 & -0.000027 & 0.0428 & 1.0000 & 0.2711 \\
MC99 & Sharpe & -0.001778 & -0.003495 & -0.000082 & 0.0422 & 1.0000 & 0.2711 \\
TC99 & Rachev 95/95 & 0.003528 & -0.010614 & 0.018498 & 0.6279 & 1.0000 & 0.7728 \\
TC99 & SS$_2$ & -0.000575 & -0.009720 & 0.005078 & 0.8830 & 1.0000 & 0.9034 \\
TC99 & STAR 95 & -0.001188 & -0.003498 & 0.000612 & 0.2567 & 1.0000 & 0.4401 \\
TC99 & Sharpe & -0.002823 & -0.007867 & 0.001073 & 0.2148 & 1.0000 & 0.4401 \\
\bottomrule\end{tabular}}
\end{table}
\FloatBarrier

\begin{table}[p]
\centering\scriptsize
\caption{Individual paired stationary-bootstrap differences, unadjusted 95\% percentile confidence intervals, and centered-bootstrap $p$-values for limited long--short portfolios, expected block length $L=40$ ($B=5{,}000$). Adjustments were calculated within the complete family of 48 comparisons.}
\label{tab:bootstrap_alt_40_LS_2026}
\setlength{\tabcolsep}{2.7pt}
\resizebox{\textwidth}{!}{%
\begin{tabular}{llrrrrrr}\toprule
Portfolio & Ratio & Difference & CI lower & CI upper & Raw $p$ & Holm $p$ & BH $p$ \\
\midrule
MV & Rachev 95/95 & -0.000298 & -0.000636 & 0.000137 & 0.1344 & 1.0000 & 0.3811 \\
MV & SS$_2$ & -0.000164 & -0.000295 & -0.000022 & 0.0180 & 0.8638 & 0.2711 \\
MV & STAR 95 & -0.000038 & -0.000068 & -0.000005 & 0.0252 & 1.0000 & 0.2711 \\
MV & Sharpe & -0.000076 & -0.000140 & -0.000007 & 0.0282 & 1.0000 & 0.2711 \\
TV & Rachev 95/95 & -0.003698 & -0.016023 & 0.009296 & 0.5445 & 1.0000 & 0.6878 \\
TV & SS$_2$ & -0.000990 & -0.009746 & 0.004900 & 0.7820 & 1.0000 & 0.8730 \\
TV & STAR 95 & -0.001120 & -0.003817 & 0.000965 & 0.3391 & 1.0000 & 0.5087 \\
TV & Sharpe & -0.002570 & -0.008489 & 0.001963 & 0.3233 & 1.0000 & 0.5044 \\
MC95 & Rachev 95/95 & -0.004211 & -0.011944 & 0.003940 & 0.2755 & 1.0000 & 0.4561 \\
MC95 & SS$_2$ & -0.002847 & -0.005935 & -0.000422 & 0.0452 & 1.0000 & 0.2711 \\
MC95 & STAR 95 & -0.000903 & -0.001978 & 0.000072 & 0.0802 & 1.0000 & 0.3448 \\
MC95 & Sharpe & -0.001852 & -0.004096 & 0.000255 & 0.0934 & 1.0000 & 0.3448 \\
TC95 & Rachev 95/95 & -0.000963 & -0.016875 & 0.015226 & 0.9034 & 1.0000 & 0.9034 \\
TC95 & SS$_2$ & -0.001464 & -0.011529 & 0.005126 & 0.7303 & 1.0000 & 0.8346 \\
TC95 & STAR 95 & -0.001385 & -0.004058 & 0.000674 & 0.2492 & 1.0000 & 0.4401 \\
TC95 & Sharpe & -0.003228 & -0.008994 & 0.001324 & 0.2138 & 1.0000 & 0.4401 \\
MC99 & Rachev 95/95 & -0.003535 & -0.012380 & 0.005999 & 0.4235 & 1.0000 & 0.5979 \\
MC99 & SS$_2$ & -0.002463 & -0.005782 & 0.000024 & 0.0884 & 1.0000 & 0.3448 \\
MC99 & STAR 95 & -0.000874 & -0.001870 & 0.000070 & 0.0766 & 1.0000 & 0.3448 \\
MC99 & Sharpe & -0.001795 & -0.003893 & 0.000135 & 0.0814 & 1.0000 & 0.3448 \\
TC99 & Rachev 95/95 & -0.002046 & -0.018400 & 0.014126 & 0.8036 & 1.0000 & 0.8767 \\
TC99 & SS$_2$ & -0.001514 & -0.011238 & 0.004878 & 0.7193 & 1.0000 & 0.8346 \\
TC99 & STAR 95 & -0.001359 & -0.003999 & 0.000679 & 0.2503 & 1.0000 & 0.4401 \\
TC99 & Sharpe & -0.003114 & -0.008783 & 0.001388 & 0.2192 & 1.0000 & 0.4401 \\
\bottomrule\end{tabular}}
\end{table}
\FloatBarrier

\section{Additional Tail Figure}\label{app:tailfig2026}
\begin{figure}[htbp]\centering\includegraphics[width=0.95\textwidth]{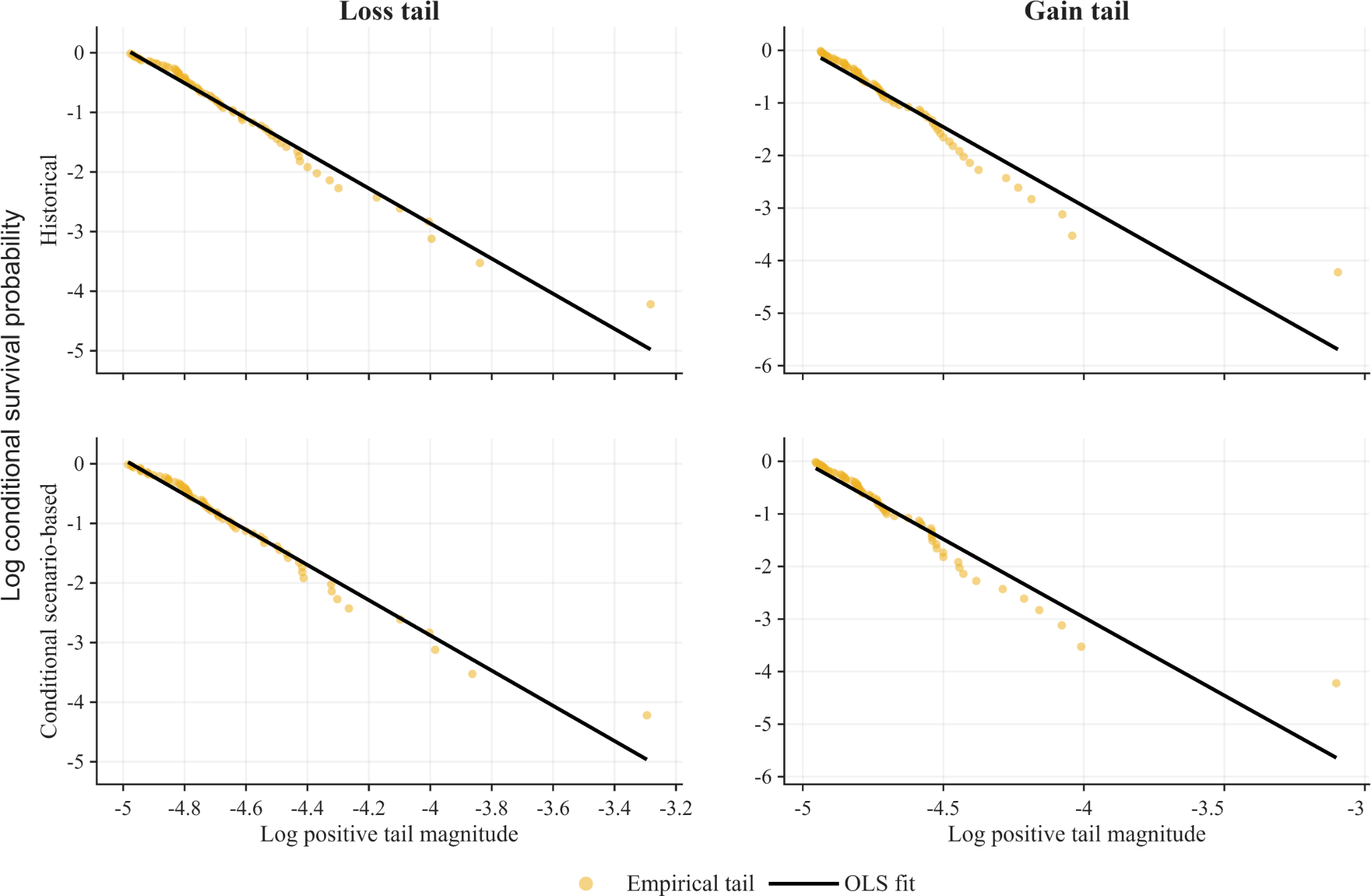}\caption{Descriptive 7\% Pareto OLS fits for LS MC99 positive loss and gain tails. Historical method top row; dynamic method bottom row.}\label{fig:paretomc992026}\end{figure}

\section{ETF allocations, concentration, and exposure}\label{app:allocation_reconciled_2026}
Table~\ref{tab:allocation_reconciled_all_2026} describes the allocations across 946 evaluated dates. Bond total is the mean signed allocation to BND and BYLD. Largest ETF is the mean of the largest absolute ETF position on each date. The Herfindahl index (HHI) uses absolute positions normalized by gross exposure and is averaged across dates. Gross and short exposures are calculated from the same dated portfolio weights. The table describes allocations to ETFs rather than the ultimate economic exposures within their underlying holdings.
\begin{table}[htbp]\centering\scriptsize
\caption{Baseline ETF weights: direct-bond allocation, concentration, and exposure. Values except HHI are percentages of portfolio capital.}\label{tab:allocation_reconciled_all_2026}
\resizebox{\textwidth}{!}{%
\begin{tabular}{lllrrrrrrr}\toprule
Method & Bounds & Objective & BND & BYLD & Bond total & Largest ETF & HHI & Gross & Short \\
\midrule
Historical & LO & MV & 3.79 & 3.62 & 7.41 & 3.79 & 0.0335 & 100.00 & 0.00 \\
Historical & LO & TV & 5.07 & 3.41 & 8.48 & 6.20 & 0.0363 & 100.00 & 0.00 \\
Historical & LO & MC95 & 8.57 & 3.33 & 11.90 & 8.57 & 0.0388 & 100.00 & 0.00 \\
Historical & LO & TC95 & 1.40 & 3.33 & 4.73 & 6.51 & 0.0368 & 100.00 & 0.00 \\
Historical & LO & MC99 & 9.56 & 3.33 & 12.90 & 9.56 & 0.0402 & 100.00 & 0.00 \\
Historical & LO & TC99 & 1.42 & 3.33 & 4.75 & 6.44 & 0.0368 & 100.00 & 0.00 \\
Historical & LS & MV & 3.36 & 3.35 & 6.71 & 3.36 & 0.0333 & 100.00 & 0.00 \\
Historical & LS & TV & 5.29 & 3.40 & 8.69 & 6.23 & 0.0350 & 103.88 & 1.94 \\
Historical & LS & MC95 & 8.65 & 3.33 & 11.98 & 8.65 & 0.0379 & 101.82 & 0.91 \\
Historical & LS & TC95 & 1.47 & 3.33 & 4.81 & 6.46 & 0.0357 & 102.96 & 1.48 \\
Historical & LS & MC99 & 9.56 & 3.33 & 12.90 & 9.56 & 0.0394 & 101.41 & 0.70 \\
Historical & LS & TC99 & 1.53 & 3.33 & 4.87 & 6.40 & 0.0357 & 102.90 & 1.45 \\
Dynamic & LO & MV & 3.77 & 3.63 & 7.40 & 3.77 & 0.0335 & 100.00 & 0.00 \\
Dynamic & LO & TV & 3.97 & 3.48 & 7.45 & 4.18 & 0.0339 & 100.00 & 0.00 \\
Dynamic & LO & MC95 & 8.37 & 3.38 & 11.76 & 8.37 & 0.0376 & 100.00 & 0.00 \\
Dynamic & LO & TC95 & 3.17 & 3.22 & 6.39 & 4.47 & 0.0337 & 100.00 & 0.00 \\
Dynamic & LO & MC99 & 8.74 & 3.35 & 12.09 & 8.74 & 0.0380 & 100.00 & 0.00 \\
Dynamic & LO & TC99 & 3.16 & 3.24 & 6.40 & 4.45 & 0.0337 & 100.00 & 0.00 \\
Dynamic & LS & MV & 3.40 & 3.36 & 6.75 & 3.40 & 0.0333 & 100.00 & 0.00 \\
Dynamic & LS & TV & 3.90 & 3.47 & 7.36 & 4.15 & 0.0339 & 100.00 & 0.00 \\
Dynamic & LS & MC95 & 8.37 & 3.38 & 11.76 & 8.37 & 0.0376 & 100.00 & 0.00 \\
Dynamic & LS & TC95 & 3.17 & 3.22 & 6.39 & 4.47 & 0.0337 & 100.00 & 0.00 \\
Dynamic & LS & MC99 & 8.74 & 3.35 & 12.09 & 8.74 & 0.0380 & 100.00 & 0.00 \\
Dynamic & LS & TC99 & 3.16 & 3.24 & 6.40 & 4.45 & 0.0337 & 100.00 & 0.00 \\
\bottomrule\end{tabular}}
\end{table}

\FloatBarrier

\section{Detailed Sensitivity Results}\label{app:sensitivity2026}
The following tables report portfolio-level estimates from four sensitivity experiments. Annualized geometric-return changes are expressed in percentage points (pp); all comparisons use matched evaluation dates. The sensitivity experiments are descriptive and do not constitute independent tests of statistical significance.
\subsection{Alternative daily target-turnover cap}
\begin{table}[htbp]\centering\scriptsize
\caption{Baseline versus 1\% daily target-turnover cap for all portfolios. Gross and net annualized geometric returns and annualized implementation turnover are percentages; changes are in percentage points. Modeled net returns assume 10 basis points per unit of one-way implementation turnover.}\label{tab:sensitivity_turnover_all}
\setlength{\tabcolsep}{3pt}
\resizebox{\textwidth}{!}{%
\begin{tabular}{lllrrrrrr}\toprule
Method & Bounds & Objective & Gross base & Gross 1\% & $\Delta$ gross & Net base & Net 1\% & Impl. TO base / 1\% \\ \midrule
Hist. & LO & MV & 11.95 & 5.18 & -6.77 & 11.90 & 5.11 & 41.5 / 73.0 \\
Hist. & LO & TV & 12.09 & 13.28 & +1.19 & 12.04 & 13.04 & 43.6 / 214.4 \\
Hist. & LO & MC95 & 11.51 & 4.79 & -6.72 & 11.47 & 4.70 & 41.5 / 86.9 \\
Hist. & LO & TC95 & 12.90 & 13.71 & +0.82 & 12.85 & 13.44 & 44.8 / 241.0 \\
Hist. & LO & MC99 & 11.60 & 5.03 & -6.57 & 11.55 & 4.93 & 41.7 / 89.3 \\
Hist. & LO & TC99 & 12.89 & 14.19 & +1.29 & 12.84 & 13.92 & 44.8 / 234.7 \\
Hist. & LS & MV & 12.09 & 3.58 & -8.50 & 12.04 & 3.45 & 41.7 / 133.7 \\
Hist. & LS & TV & 12.25 & 16.19 & +3.94 & 12.20 & 15.80 & 46.1 / 340.0 \\
Hist. & LS & MC95 & 11.62 & 3.82 & -7.81 & 11.58 & 3.59 & 42.7 / 217.6 \\
Hist. & LS & TC95 & 12.93 & 17.06 & +4.12 & 12.88 & 16.65 & 46.2 / 348.0 \\
Hist. & LS & MC99 & 11.61 & 4.14 & -7.47 & 11.57 & 3.92 & 42.6 / 215.2 \\
Hist. & LS & TC99 & 12.92 & 17.06 & +4.13 & 12.87 & 16.65 & 46.1 / 346.1 \\
Dyn. & LO & MV & 11.86 & 5.10 & -6.76 & 11.81 & 4.91 & 41.5 / 183.6 \\
Dyn. & LO & TV & 11.98 & 13.11 & +1.12 & 11.94 & 12.80 & 42.3 / 275.3 \\
Dyn. & LO & MC95 & 11.33 & 5.20 & -6.13 & 11.28 & 4.92 & 41.5 / 264.8 \\
Dyn. & LO & TC95 & 12.35 & 14.05 & +1.70 & 12.30 & 13.73 & 43.4 / 277.7 \\
Dyn. & LO & MC99 & 11.32 & 4.87 & -6.44 & 11.27 & 4.60 & 41.5 / 264.9 \\
Dyn. & LO & TC99 & 12.35 & 14.66 & +2.32 & 12.30 & 14.35 & 43.4 / 277.9 \\
Dyn. & LS & MV & 12.07 & 3.17 & -8.90 & 12.02 & 2.98 & 41.7 / 189.6 \\
Dyn. & LS & TV & 12.00 & 13.73 & +1.73 & 11.95 & 13.38 & 42.2 / 314.3 \\
Dyn. & LS & MC95 & 11.33 & 3.43 & -7.90 & 11.28 & 3.10 & 41.5 / 317.5 \\
Dyn. & LS & TC95 & 12.35 & 15.03 & +2.68 & 12.30 & 14.65 & 43.4 / 325.7 \\
Dyn. & LS & MC99 & 11.32 & 3.07 & -8.24 & 11.27 & 2.75 & 41.5 / 318.9 \\
Dyn. & LS & TC99 & 12.35 & 15.19 & +2.84 & 12.30 & 14.82 & 43.4 / 323.9 \\
\bottomrule\end{tabular}}\end{table}\FloatBarrier
\subsection{Scenario count and simulation seed}
\begin{table}[htbp]\centering\small
\caption{Changes in annualized geometric returns (pp) for dynamic portfolios relative to the 10,000-scenario baseline. Simulation seed and other settings are held fixed.}\label{tab:sensitivity_scenarios_all}
\begin{tabular}{llrr}\toprule Portfolio & Objective & 5,000 minus 10,000 & 20,000 minus 10,000 \\ \midrule
LO & MV & -0.000 & +0.000 \\
LO & TV & +0.008 & -0.013 \\
LO & MC95 & +0.016 & +0.007 \\
LO & TC95 & +0.007 & -0.003 \\
LO & MC99 & +0.017 & +0.011 \\
LO & TC99 & +0.008 & -0.002 \\
LS & MV & -0.001 & -0.001 \\
LS & TV & +0.002 & -0.023 \\
LS & MC95 & +0.016 & +0.007 \\
LS & TC95 & +0.007 & -0.003 \\
LS & MC99 & +0.017 & +0.011 \\
LS & TC99 & +0.008 & -0.002 \\
\bottomrule\end{tabular}\end{table}
\begin{table}[htbp]\centering\small
\caption{Effect of changing the dynamic simulation seed from 20261015 to 12345: change in annualized geometric returns (pp). Historical returns are unchanged under the alternative seed.}\label{tab:sensitivity_seed_all}
\begin{tabular}{llr}\toprule Bounds & Objective & Dynamic seed change \\ \midrule
LO & MV & -0.001 \\
LO & TV & -0.026 \\
LO & MC95 & +0.012 \\
LO & TC95 & -0.007 \\
LO & MC99 & +0.000 \\
LO & TC99 & -0.003 \\
LS & MV & -0.002 \\
LS & TV & -0.019 \\
LS & MC95 & +0.012 \\
LS & TC95 & -0.007 \\
LS & MC99 & +0.000 \\
LS & TC99 & -0.003 \\
\bottomrule\end{tabular}\end{table}\FloatBarrier
\subsection{Estimation-window length}
\begin{table}[htbp]\centering\scriptsize
\caption{Matched annualized geometric-return differences (pp) from 1,008 observations for alternative moving estimation windows. All rows use the same 694 out-of-sample dates (22 December 2023--30 September 2026).}\label{tab:sensitivity_window_all}
\begin{tabular}{lllrr}\toprule Method & Bounds & Objective & 756 minus 1008 & 1260 minus 1008 \\ \midrule
Hist. & LO & MV & -0.034 & +0.107 \\
Hist. & LO & TV & +0.022 & +0.270 \\
Hist. & LO & MC95 & -0.456 & +0.424 \\
Hist. & LO & TC95 & -0.302 & -0.346 \\
Hist. & LO & MC99 & -0.431 & +0.250 \\
Hist. & LO & TC99 & -0.233 & -0.473 \\
Hist. & LS & MV & -0.012 & +0.010 \\
Hist. & LS & TV & -0.070 & +0.277 \\
Hist. & LS & MC95 & -0.407 & +0.294 \\
Hist. & LS & TC95 & -0.226 & -0.362 \\
Hist. & LS & MC99 & -0.524 & +0.184 \\
Hist. & LS & TC99 & -0.196 & -0.462 \\
Dyn. & LO & MV & -0.346 & +0.150 \\
Dyn. & LO & TV & -0.156 & -0.058 \\
Dyn. & LO & MC95 & -0.377 & +0.322 \\
Dyn. & LO & TC95 & -0.011 & -0.158 \\
Dyn. & LO & MC99 & -0.363 & +0.323 \\
Dyn. & LO & TC99 & -0.007 & -0.163 \\
Dyn. & LS & MV & -0.057 & +0.008 \\
Dyn. & LS & TV & -0.109 & -0.034 \\
Dyn. & LS & MC95 & -0.368 & +0.322 \\
Dyn. & LS & TC95 & -0.011 & -0.158 \\
Dyn. & LS & MC99 & -0.363 & +0.323 \\
Dyn. & LS & TC99 & -0.007 & -0.163 \\
\bottomrule\end{tabular}\end{table}\FloatBarrier

\FloatBarrier
\begin{adjustwidth}{-\extralength}{0cm}
\reftitle{References}
\bibliography{references}
\end{adjustwidth}
\PublishersNote{}
\end{document}